\documentstyle[epsfig,psfig]{mn}

\input{epsf}

\begin{document}

\title [The QSO Power Spectrum] { The 2dF QSO Redshift Survey - XI. The
  QSO Power Spectrum }

\author[P.~J. Outram et~al. ] { P.~J. Outram$^1$, Fiona Hoyle$^{2}$, T.
  Shanks$^1$, S. M. Croom$^{3}$, B.~J. Boyle$^{3}$, \newauthor L.
  Miller$^{4}$, R.~J. Smith$^{5}$ \& A.~D.~Myers$^1$
  \\
  1 Department of Physics, Science Laboratories, South Road, Durham,
  DH1 3LE, U.K.
  \\
  2 Department of Physics, Drexel University, 3141 Chestnut Street, Philadelphia, PA 19104, U.S.A.\\
  3 Anglo-Australian Observatory, PO Box 296, Epping, NSW 2121, Australia\\
  4 Department of Physics, Oxford University, Keble Road, Oxford, OX1 3RH, U.K.\\
  5 Liverpool John Moores University, Twelve Quays House, Egerton Wharf, Birkenhead, CH41 1LD, U.K. \\
}

\maketitle
 
\begin{abstract}
  
  We present a power spectrum analysis of the final 2dF QSO Redshift
  Survey catalogue containing 22652 QSOs. Utilising the huge volume
  probed by the QSOs, we can accurately measure power out to scales of
  $\sim 500h^{-1}$Mpc and derive new constraints, at $z\sim1.4$, on the
  matter and baryonic contents of the Universe. Importantly, these new
  cosmological constraints are derived at an intermediate epoch between
  the CMB observations at $z\sim1000$, and local ($z\sim0$) studies of
  large-scale structure; the average QSO redshift corresponds to a
  look-back time of approximately two-thirds of the age of the
  Universe. We find that the amplitude of clustering of the QSOs at
  $z\sim1.4$ is similar to that of present day galaxies. The power
  spectra of the QSOs at high and low redshift are compared and we find
  little evidence for any evolution in the amplitude. Assuming a
  $\Lambda$ cosmology to derive the comoving distances, $r(z)$, to the
  QSOs, the power spectrum derived can be well described by a model
  with shape parameter $\Gamma=0.13\pm0.02$. If an Einstein-de Sitter
  model $r(z)$ is instead assumed, a slightly higher value of
  $\Gamma=0.16\pm0.03$ is obtained. A comparison with the {\it Hubble
    Volume} $\Lambda$CDM simulation shows very good agreement over the
  whole range of scales considered. A standard ($\Omega_m=1$) CDM
  model, however, predicts a much higher value of $\Gamma$ than is
  observed, and it is difficult to reconcile such a model with these
  data. We fit CDM model power spectra (assuming scale-invariant
  initial fluctuations), convolved with the survey window function, and
  corrected for redshift space distortions, and find that models with
  baryon oscillations are slightly preferred, with the baryon fraction
  $\Omega_b/\Omega_m=0.18\pm0.10$. The overall shape of the power
  spectrum provides a strong constraint on $\Omega_mh$ (where $h$ is
  the Hubble parameter), with $\Omega_mh=0.19\pm0.05$.

\end{abstract}


\section{Introduction}

The large-scale structure of the Universe represents one of the most
powerful discriminants between cosmological models.  QSOs are highly
effective probes of the structure of the Universe over a wide range of
scales and can trace clustering evolution over a look-back time which
is 70-80 per cent of its present age. In the linear regime ($10\la
r\la1000$h$^{-1}$Mpc), they are clearly superior to galaxies as probes
of large scale structure by virtue of both the large volumes they
sample and their flat $n(z)$ distribution. Accurately measuring the
clustering over these scales bridges the gap between the clustering
results on relatively small scales from galaxy redshift surveys out to
scales previously only probed by microwave background anisotropy
experiments.

The power spectrum perhaps provides the most natural description of the
matter fluctuations that comprise large-scale structure; for a Gaussian
field, the amplitude of the Fourier modes provide a statistically
complete description of the density perturbations.  One potential
problem of using QSOs as probes of large scale structure is that they,
like galaxies, are biased tracers of the mass density field. QSO
imaging experiments at redshifts out to $z\sim 1-2$ suggest that the
QSOs reside in relatively modest ($\sim L_*$) galaxy hosts (Rix et al.
2001; Ridgway et al. 2001) and that the clustering environment of UVX
QSOs are similar to optically selected galaxies (Croom \& Shanks 1999).
This implies that at $z\la 1-2$ QSOs and galaxies have a similar bias
with respect to the underlying mass density field (Kauffmann \&
Haehnelt 2002). It is likely that the bias is scale-dependent, at least
at small scales (Blanton et al. 1999). At larger scales
($r>>10$h$^{-1}$Mpc), however, where QSO clustering statistics prevail,
the shape (although not the amplitude) of the measured power spectrum
can reasonably be assumed to apply to the mass and galaxies as well,
since it is unlikely that the QSO bias will be scale dependent in this
regime (Coles 1993; Mann, Peacock \& Heavens 1998). For the purposes of
modelling the QSO P(k) in this paper we assume that any bias is scale
independent on the scales considered ($40<r<500$h$^{-1}$Mpc).

With the principal aim of determining the form of QSO clustering, and
hence producing strong new constraints on cosmological models, we have
used the 2dF multi-fibre spectrograph on the Anglo-Australian Telescope
to make a large QSO redshift survey. The 2dF QSO Redshift Survey (2QZ)
comprises two $5\deg \times 75\deg$ declination strips, one at the
South Galactic Pole and one in an equatorial region in the North
Galactic Cap. QSOs are selected by ultra-violet excess (UVX) in the
$u-b_{\rm J}:b_{\rm J}-r$ plane.  The effectiveness of our QSO
selection criteria has been demonstrated with an overall QSO sky
density of $35 \deg^{-2}$ to $b_{\rm J}<20.85$ being achieved, at a
completeness of 93 per cent. Spectroscopic observations are now
complete, and we have been able to identify 86 per cent of the
colour-selected candidates. Around half, 47 per cent, of the 47768
candidates are positively identified as QSOs, leading to a final
catalogue containing 22652 QSOs. The QSOs typically have redshifts
$0.3<z<3$, and probe a volume of around $2\times 10^9\,h^{-3}\,$Mpc$^3$
(assuming an Einstein-de Sitter cosmology).

The spectra of the first 10000 2QZ QSOs were released in April 2001
(Croom et al 2001a) and can be obtained at {\tt
  http://www.2dfquasar.org}.  Hoyle et al. (2002) used these QSOs to
measure the QSO power spectrum.  They found that the QSO power spectrum
has similar amplitude to that of local galaxies, and little evolution
is seen with redshift. The main conclusion was that they detected
large-scale power significantly in excess of the standard
($\Omega_m=1$) CDM prediction, strongly ruling out this model and more
consistent with CDM models with lower mass content and a non-zero
cosmological constant ($\Lambda$CDM).

In this paper, we re-apply the power spectrum analysis described in
Hoyle et al. (2002) to the final 2QZ catalogue containing 22652 QSOs.
As well as having more than twice as many QSOs as available to Hoyle et
al., this dataset has a much cleaner window function, significantly
reducing the potential systematics on large scales. We briefly describe
the 2QZ data set in Section \ref{sec:qsodata}, including the angular
and radial selection functions which are required to estimate the
survey window function. Section~\ref{pkest} discusses the power
spectrum estimation method, and the resulting QSO power spectrum is
presented in Section~\ref{sec:qsoreslts}.  In this section we compare
the QSO power spectrum to that of present day galaxies and clusters,
and investigate evolution in the clustering amplitude. Then, in Section
\ref{sec:qsoimp}, we compare the QSO power spectrum with models of
large scale structure, deriving strong constraints on the baryon and
matter content of the Universe, before drawing conclusions in Section
\ref{sec:qsoconc}.

\section{The 2QZ catalogue}
\label{sec:qsodata}

The 2QZ comprises two $5\deg \times 75\deg$ declination strips, one at
the South Galactic Pole (centred at $\delta = -30^{\circ}$, with
21$^{\rm h}$40$^{\rm m}$ $\la \alpha \la $ 03$^{\rm h}$15$^{\rm m}$)
and one in an equatorial region in the North Galactic Cap (centred at
$\delta= 0^{\circ}$ with 09$^{\rm h}$50$^{\rm m}$ $\la \alpha \la $
14$^{\rm h}$50$^{\rm m}$).  We will refer to these regions as the SGC
and NGC respectively.  QSOs are selected to $b_{\rm J}<20.85$ by
ultra-violet excess (UVX) in the $u-b_{\rm J}:b_{\rm J}-r$ plane, at a
completeness of 93 per cent.  The spectra of the objects are obtained
using the 2dF instrument on the AAT and reduced using the 2dF pipeline
reduction system (Bailey \& Glazebrook 1999). Objects are identified as
QSOs by an automated procedure known as {\tt AUTOZ} (Miller et al., in
prep.). {\tt AUTOZ} also determines the QSO redshifts; these have then
been visually checked by two independent observers.  The final 2QZ
catalogue, containing 22652 QSOs (only those QSOs with quality 1 are
used in this analysis; see Croom et al. (2001a) for details), covers an
area of 740 deg$^2$, and the data will be released to the community in
the first half of 2003.

\subsection{The QSO selection function}
\label{sec:rancat}

In order to estimate the QSO power spectrum, we need to take into
account the various selection effects introduced when constructing the
survey. This is achieved by generating a catalogue of random points
that mimics the angular and radial selection functions of the QSOs but
otherwise is unclustered.

The rapid luminosity evolution seen in QSOs (Boyle et al. 2000) means
that at each redshift, over the redshift range we are considering, we
are studying roughly the same part of the QSO luminosity function,
relative to $M^*$. In areas where the spectroscopic completeness is
low, due for example to poor seeing conditions, we are sampling on
average slightly brighter QSOs. This just has the effect of sampling
QSOs further up the luminosity function at all redshifts, rather than
sampling objects at lower redshift, as is typical with galaxy redshift
surveys. Even if we cut back the magnitude limit of the survey by half
a magnitude from $b_{\rm J}=20.85$ to $b_{\rm J}=20.35$, the mean
redshift only drops slightly from $\bar z = 1.48$ to $\bar z = 1.45$.
Therefore, for this work we have assumed that the radial and angular
selection functions can be applied independently, and consider each in
turn below.

\subsubsection{The angular selection function}
\label{sec:rancat_ang}

The QSOs were observed simultaneously with galaxies from the 2dF Galaxy
Redshift Survey (2dFGRS), and an adaptive tiling algorithm was
developed to allow as many galaxies and QSOs as possible to be observed
in each pointing. The pointings overlap in high density regions to
maximise the coverage in all areas of the survey, and with this
algorithm, we have achieved a completeness of 93 per cent. In addition
to the 7 per cent of objects on which we were unable to place fibres,
regions around bright stars were omitted from the input catalogue, and
hence the angular mask of the 2QZ is quite complicated.

The survey also suffers from spectroscopic incompleteness. We have
failed to confirm the identity of approximately 14 per cent of the
objects observed. The spectra of these objects are typically very
noisy, with a lack of discernible spectral features. As some classes of
objects are easier to identify in low S/N spectra than others, due to,
for example, distinctive emission lines, the spectroscopic completeness
of QSOs in the survey may differ from that of stars, or the input
catalogue as a whole. To correct for spectroscopic incompleteness we
therefore need to estimate the fraction of QSOs that remain in these
unidentified spectra. By examining the results from objects that have
been observed more than once, such an estimate can be obtained. For
objects unidentified after their first observation, but identified in
subsequent observations, the relative fractions of QSOs and stars are
remarkably similar to those of the identified objects in the full
survey. Hence we estimate that QSOs account for approximately the same
proportion ($\sim59$ per cent) of the identified and unidentified
populations and so the spectroscopic incompleteness of QSOs is in fact
the same as that of the input catalogue as a whole. However, the rate
of identifications obtained through the re-observation of previously
unidentified objects is significantly lower than the survey average,
and hence it is still possible that the fraction of QSOs remaining in
the unidentified population could be somewhat lower. Through
consideration of the fraction of QSOs unidentified after their first
observation that are subsequently identified, we can set a lower limit
to the fraction of QSOs in the remaining unidentified population at
about 40 per cent. We will consider the effect of different
spectroscopic completeness corrections in Section~\ref{pksys}.

To match the angular selection function of the QSOs, we construct a
completeness map, estimating completeness both to observational and
spectroscopic effects in regions defined by the intersection of 2dF
pointings. In each such region we calculate the fraction of objects
contained in the input catalogue that have been identified. As the
spectroscopic incompleteness of QSOs is approximately the same as that
of the full input catalogue, the proportion of QSOs in unobserved
objects should be the same as the proportion of QSOs in observed, but
unidentified objects. Hence we can take this fraction to be the overall
(observational and spectroscopic) completeness in that region. As we
are applying a spectroscopic completeness correction, we have decided
not to apply simultaneously a minimum spectroscopic completeness cut,
as this would reduce the size of our sample.

We also account for the extinction due to Galactic dust, using the
estimate of the dust reddening, $E(B-V)$, as a function of position
given by Schlegel, Finkbeiner \& Davis (1998).  As Galactic dust
changes the effective magnitude limit, we weight the random
distribution by
\begin{equation}
W_{\rm ext}(\alpha,\delta) = 10^{- \beta {\rm A_{\rm b_J}}}
\end{equation}
where ${\rm A_{b_J}} = 4.035E(B-V)$ and $\beta=0.3$, the slope of the
integral QSO number-magnitude relation at the magnitude limit of the
survey, $b_{\rm J}$=20.85. As we apply a dust correction to the angular
mask, we need to correct for dust extinction in the observed radial
selection function (below) to avoid including the effect of dust twice.

\subsubsection{The radial selection function}

We are measuring QSO clustering as a function of comoving distance, and
so we need to adopt a cosmology to convert from redshift into comoving
distance. In this paper we consider an Einstein-de Sitter ($\Omega_{\rm
  m}$=1.0, $\Omega_{\Lambda}$=0.0) cosmology $r(z)$ (EdS hereafter) and
an $\Omega_{\rm m}$=0.3, $\Omega_{\Lambda}$=0.7 cosmology $r(z)$
($\Lambda$ hereafter).

\begin{figure} 
  \centerline{\hbox{\psfig{figure=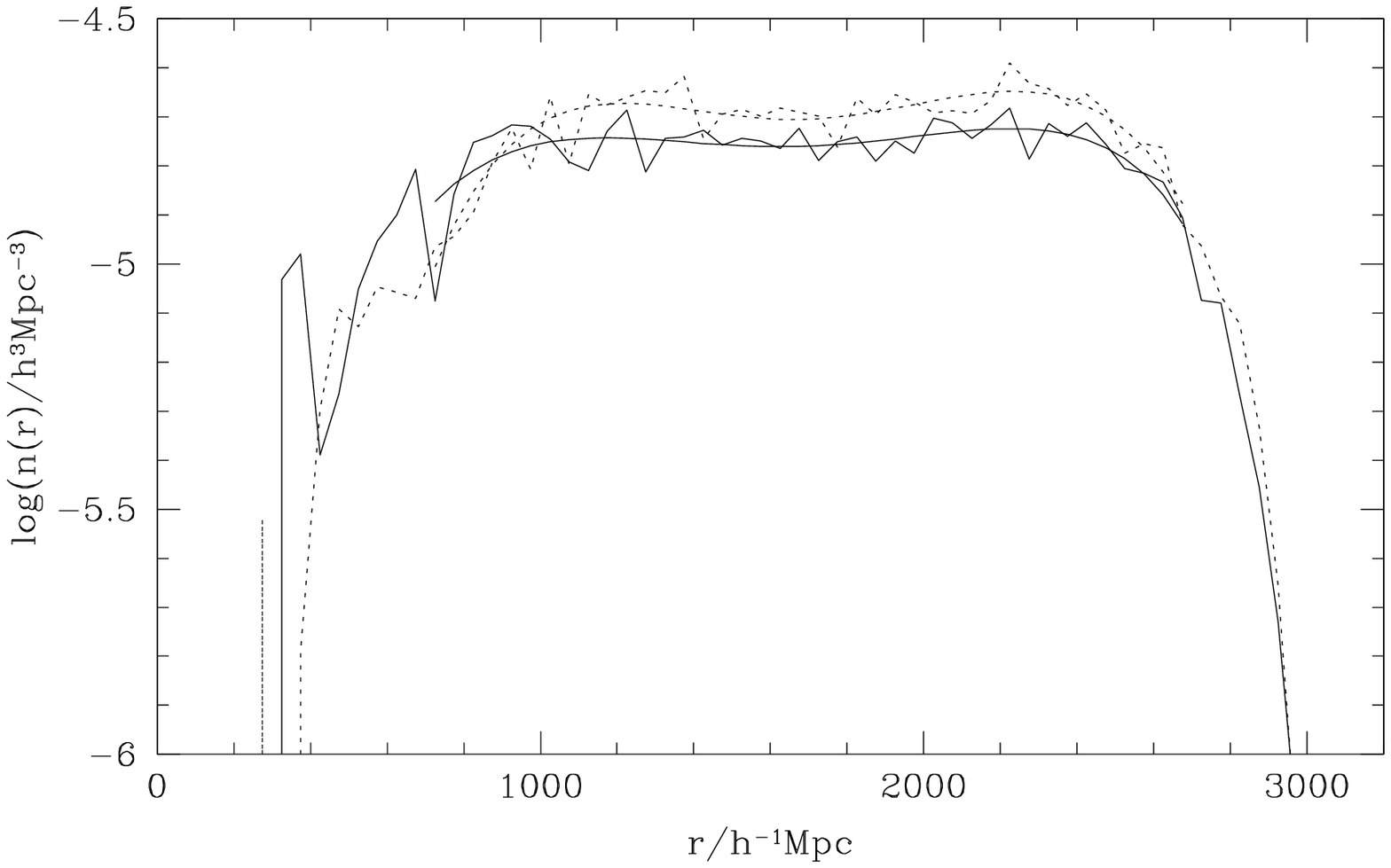,width=8.5cm}}}
  \centerline{\hbox{\psfig{figure=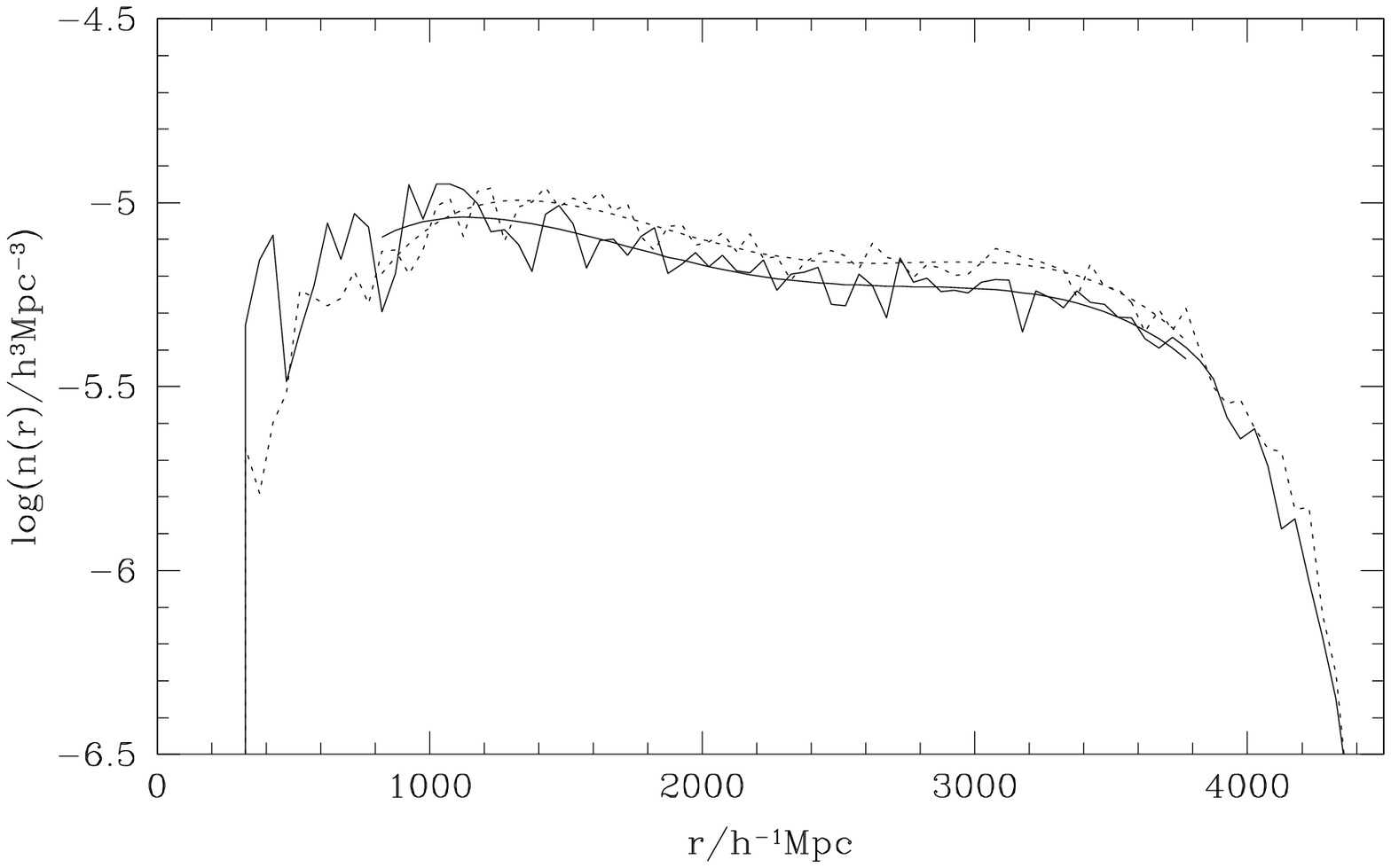,width=8.5cm}}}
\caption
{The radial number density of QSOs in bins of $\Delta r = 50
  h^{-1}$Mpc in the NGC (solid line) and SGC (dashed line), corrected
  for the average dust extinction in each region and calculated
  assuming an EdS $r(z)$ (top) and the $\Lambda$ $r(z)$ (bottom).
  Overlaid are the 10th order polynomial fits used to create the random
  catalogues, plotted over the redshift range $0.3<z<2.2$.  The number
  density varies very slowly as a function of scale; almost constant,
  assuming the EdS $r(z)$ and decreasing only by a factor of
  approximately three over the range $0.3<z<2.2$, assuming the
  $\Lambda$ $r(z)$. At higher redshifts, the number density decreases
  rapidly.}
\label{fig:nr}
\end{figure}

Figure \ref{fig:nr} shows the radial selection function of QSOs in the
NGC (solid line) and SGC (dashed line), corrected for the average dust
extinction in each region and calculated assuming the EdS $r(z)$ or the
$\Lambda$ $r(z)$.  Due to the contribution from the host galaxy, QSOs
at low redshift are lost from our input catalogue because they do not
appear stellar. The host galaxy also reddens faint QSOs causing then to
be preferentially lost from the sample. To limit this incompleteness,
we restrict our analysis to $z>0.3$ ($r>740 h^{-1}$Mpc, assuming EdS,
or $r>835 h^{-1}$Mpc, assuming $\Lambda$).  At high redshifts the UVX
colour selection technique breaks down as the Lyman-alpha forest enters
the $U$ band, and the completeness of the survey rapidly drops.
Therefore, we also limit our analysis to $z<2.2$ ($r<2645 h^{-1}$Mpc,
assuming EdS, or $r<3820 h^{-1}$Mpc, assuming $\Lambda$). Over the
redshift range $0.3<z<2.2$ the completeness of the 2QZ colour selection
technique is $\ga90$ per cent.  The SGC has a slightly higher
normalisation (15 per cent) than the NGC, even after correcting for the
higher levels of dust in the NGC, but very similar shape, except at low
redshift ($r<1200 h^{-1}$Mpc, or $z<0.45$) where the SGC suffers from
higher incompleteness, probably due to star/galaxy separation
differences. For this reason we treat the radial selection function
separately in the two strips. The QSO number density varies very slowly
as a function of redshift. Assuming EdS it is almost constant over the
range $0.3<z<2.2$, whereas assuming $\Lambda$, it decreases only by a
factor of approximately three over the range $0.3<z<2.2$. Applying
these redshift cuts therefore allows us to consider clustering
statistics without the need for a weighting scheme.  This gives us a
sample of 8704 QSOs in the NGC, and 10845 QSOs in the SGC.

We estimate the space density of QSOs as a function of comoving
distance, $n(r)$, via a 10 coefficient (9th order) polynomial fit to the number
distribution of QSOs, $N(r)$, in each strip in bins of $\Delta r = 50
h^{-1}$Mpc. Further details of the fits, including the coefficients,
are given in Table~\ref{tablefit}.  The resulting $n(r)$ distributions,
used to create the random catalogue, are plotted in Fig.~\ref{fig:nr}
over the redshift range $0.3<z<2.2$. A low order polynomial would
provide a poor fit to the $n(r)$ distribution. As the order is
increased, the quality of the fit, as measured by the $\chi^2$
statistic, improves dramatically until approximately an 8/9 coefficient
polynomial is used. Further increasing the order of the fit, however,
hardly reduces the $\chi^2$ value. If a much higher order polynomial
were used, it may also start to fit real features in the $n(r)$
distribution, and hence artificially remove power. Hence a 10 coefficient
polynomial is chosen so as to provide an adequate fit to the $n(r)$
distribution without over-fitting.

\begin{table}

\caption{The radial selection function is fitted separately in NGC and
  SGC, assuming either the EdS $r(z)$ or the $\Lambda$ $r(z)$ to derive
  comoving distances from the QSO redshifts. This table shows the
  values of the 10 coefficients obtained from a 9th order polynomial fit,
  $N(r)=\sum_{i=0}^9 c_{_i} (\frac{r}{1000})^i$, to the number of QSOs in each strip,
  in the range $500<r<4200 h^{-1}$Mpc ($\Lambda$), or $500<r<2800
  h^{-1}$Mpc (EdS), in bins of $\Delta r = 50h^{-1}$Mpc. The QSO number
  density, $n(r)$, is then obtained via $n(r)=\frac{a}{r^2}N(r)$, where
  $a_{_{NGC}}=0.2456$, and $a_{_{SGC}}=0.2281$ are constants that take
  into account the average completeness and extinction in each strip. 
}
\begin{center}
\begin{tabular}{@{}crrrr}
\hline
&NGC, EdS&NGC, $\Lambda$&SGC, EdS&SGC, $\Lambda$\\
\hline
$c_{_0}$&104.623&47.1019&295.615&124.692\\
$c_{_1}$&-448.886&-216.212&-987.737&-440.183\\
$c_{_2}$&576.207&339.671&615.208&463.583\\
$c_{_3}$&-35.6319&-154.792&883.628&-44.2143\\
$c_{_4}$&-176.484&18.5957&-858.731&-92.7632\\
$c_{_5}$&17.0201&0.628647&-20.3508&20.7977\\
$c_{_6}$&48.8416&1.83275&188.357&8.12686\\
$c_{_7}$&-13.6702&-0.301663&-13.6225&-2.35053\\
$c_{_8}$&-0.639615&-0.126688&-21.5386&-0.123221\\
$c_{_9}$&0.269846&0.022045&4.13687&0.049514\\
\hline
\end{tabular}
\end{center}

\label{tablefit}
\end{table}

A poor fit to the $n(r)$ distribution would artificially introduce
waves in the normalised QSO distribution, and hence add power on the
largest scales. Indeed, when the $\Lambda$ power spectrum is measured
using a radial selection function derived from a 5 coefficient polynomial
fit, and comparing to the result using a 10 coefficient polynomial fit, we find
that the power is significantly enhanced on scales $>200h^{-1}$Mpc
($\log (k / h \,{\rm Mpc}^{-1} ) < -1.5 $).  Slightly increasing the
order of the polynomial fit, however, has little effect our power
spectrum estimate. Remeasuring the $\Lambda$ power spectrum using a
15 coefficient  polynomial fit, and comparing to the result using a 10 coefficient 
polynomial fit, we find that out to scales of 400$h^{-1}$Mpc ($\log
(k / h \,{\rm Mpc}^{-1} ) > -1.8 $), the power spectra agree in
amplitude to within 4 per cent, considerably smaller than the
fractional error at this scale.

\subsection{Mock 2QZ catalogues}
\label{hubvol}

To test our method of power spectrum estimation described below, as
well as our error estimates, we have used the huge {\it Hubble Volume}
$\Lambda$CDM simulation (Frenk et al. 2000, Evrard et al. 2002), where
the particles have been output along an observer's past light cone to
simulate the 2QZ. The cosmological parameters of the simulation are
$\Omega_{\rm b}$=0.04, $\Omega_{\rm CDM}$=0.26, $\Omega_{\Lambda}$=0.7,
$H_{0}$=70 km s$^{-1}$Mpc$^{-1}$ and $\sigma_8=0.9$. One billion mass
particles are contained within a cube that is 3,000$h^{-1}$Mpc on a
side. To create realistic QSO mock catalogues to match the final 2QZ
sample, we bias the mass particles to give a similar clustering
pattern, and introduce the same angular and radial selection function
as seen in the final 2QZ sample, described in the previous section. For
further details see Hoyle (2000), where the {\it Hubble Volume}
simulation was compared to the 10k QSO catalogue.

\section{Power spectrum estimation}
\label{pkest}
The power spectrum estimation is carried out as described in Hoyle et
al. (2002), using the method outlined in Tadros \& Efstathiou (1996).
The two declination strips are treated separately. They are embedded
into a larger cubical volume, and the density field is binned onto a
256$^3$ mesh, using nearest grid-point assignment.  The power spectrum
of each region is estimated using a Fast Fourier Transform (FFT), and
the average of the resulting power spectra is taken. The Fourier modes
are binned up logarithmically to reduce the covariance between each
bin.  The FFT is unreliable at small scales; for a 256$^3$ FFT and a
box of size 4000$h^{-1}$Mpc, the Nyquist frequency of the transform is
$k_{\rm Nyq}=0.2\,h \,{\rm Mpc}^{-1}$. The limit out to which the power
spectrum measurement is reliable depends on the grid assignment scheme,
and is a fraction of $k_{\rm Nyq}$. Using the nearest grid point
assignment scheme, $k_{\rm lim} \approx 1/2$ $k_{\rm Nyq}$ (Hatton
1999), which, for the above box size, corresponds to a value of $\log
(k_{\rm lim}/h \,{\rm Mpc}^{-1}) \approx -1$ or a scale of $\approx 60
h^{-1}$Mpc.

The measured power spectrum is convolved with the power spectrum of the
window function. As the QSO co-moving number density is almost constant
out to very large scales, we are effectively considering a volume
limited sample. Hence each QSO carries equal weight, and the survey
window function, $W({\mathbf x})$, simply takes a value of unity in the
volume of the universe included in the survey, and zero elsewhere. This
is approximated using a catalogue containing a large number of
unclustered points with the same radial and angular distribution as the
survey, and its power spectrum is calculated in a similar manner to
that of the data.

To enable measurements of the power spectrum on smaller scales, we use
a direct method of computing the Fourier transform.  This method is
time intensive and can only be applied to a small number of particles
(so cannot be applied to the random catalogue). Therefore, we make the
assumption that the window function has a small amplitude, and so
negligible effect on the power spectrum estimate.  Using mock
catalogues drawn from the {\it Hubble Volume} simulation, Hoyle (2000)
demonstrated that on scales smaller than $\approx 100 h^{-1}$Mpc ($\log
(k/h \,{\rm Mpc}^{-1}) \approx -1.2$) this approximation is accurate.

Finally, to obtain a single 2QZ power spectrum for each assumed
cosmology, we average the power spectrum of the QSOs in the NGC and SGC
strips together, weighting by the inverse of the variance on each
scale.

\subsection{The window function}
\label{wfsec}

\begin{figure} 
  \vspace{-2cm} \centerline{\hbox{\psfig{figure=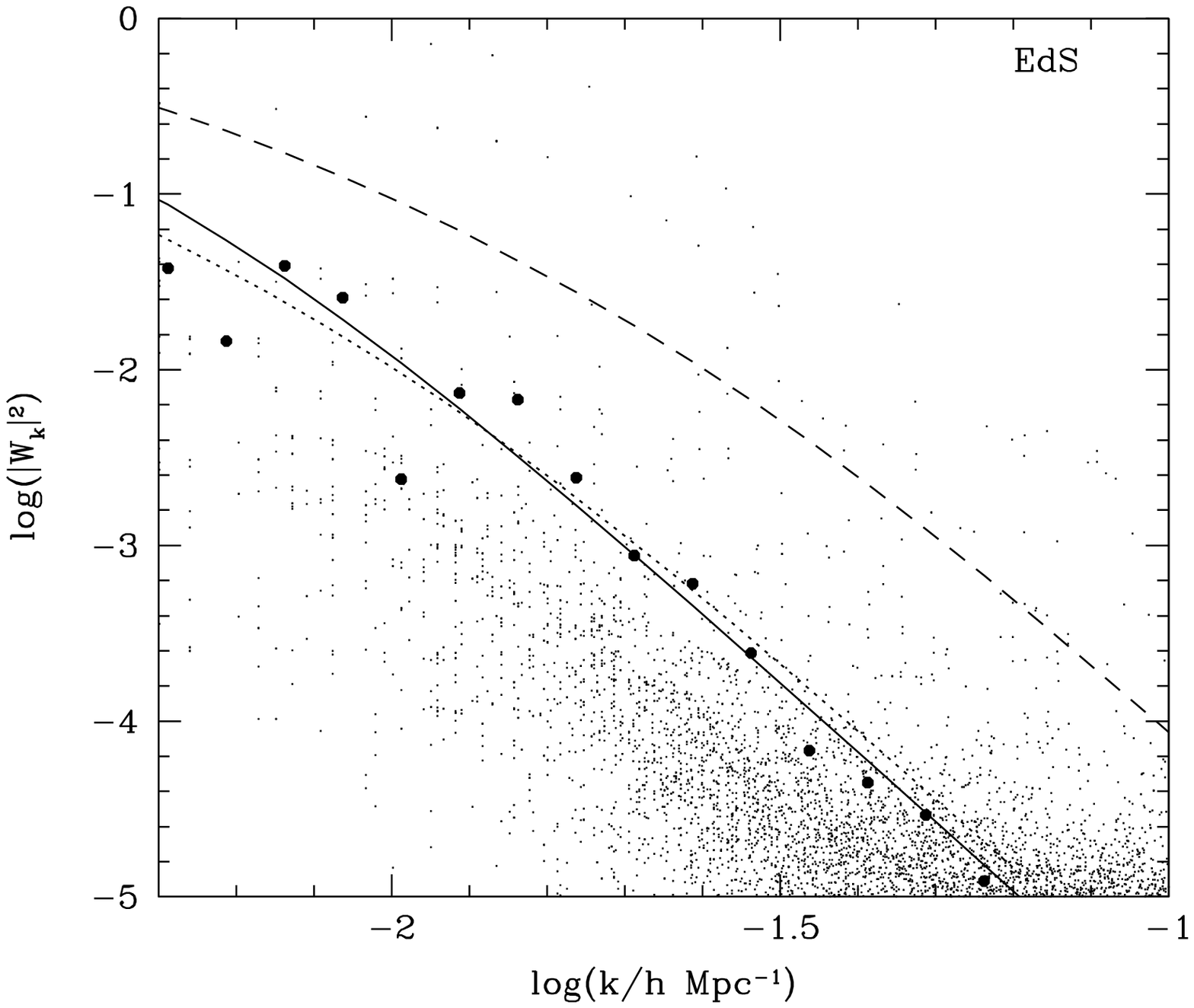,width=8.5cm}}}
  \vspace{-2cm} \centerline{\hbox{\psfig{figure=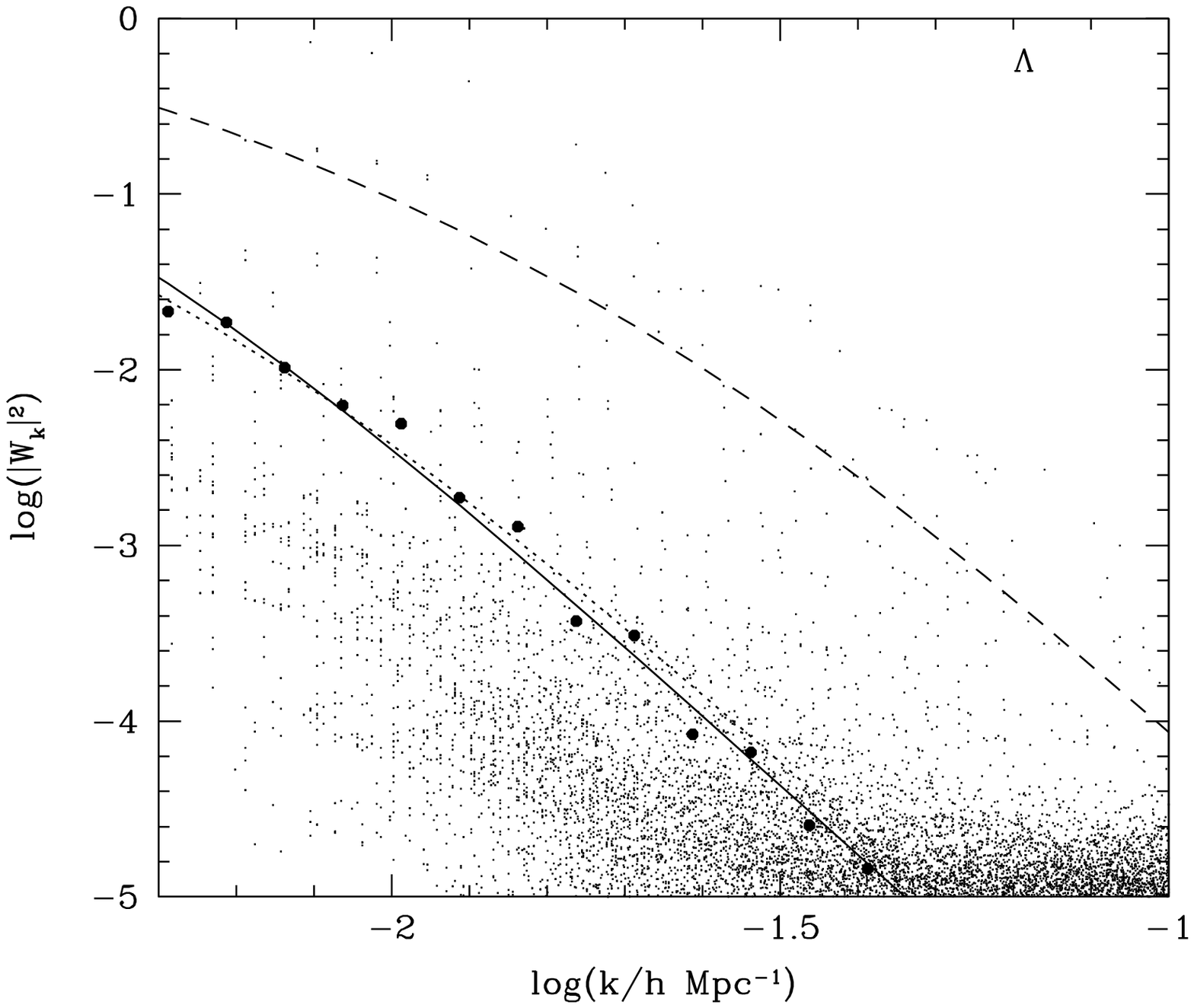,width=8.5cm}}}
\caption
{The window function of the 2QZ NGC region, measured assuming EdS (top)
  and $\Lambda$ (bottom). The window function is estimated using a
  random catalogue created to match the 2QZ selection function,
  containing 25 times as many points as the QSO catalogue. The points
  show measurements in individual $\mathbf{k}$ modes from the FFT grid.
  The tail to high k at log($|W_k|^2$)$\sim-5$ is due to shot noise in
  the window function estimate. The bold points show the binned,
  spherically averaged window function, and the solid line is an
  analytic fit to the spherically averaged window function. In either
  cosmology, the window function for the SGC is almost identical to
  that of the NGC. Due to the huge volume probed by 2QZ, the window
  function is much more compact than that of galaxy surveys, and has
  the form of a steep power law out to scales of 600$h^{-1}$Mpc. For
  comparison an analytic fit to the spherically averaged 2dFGRS window
  function is also shown (dashed line), together with the same window
  function shifted to larger scales by a factor $2.5\times$ (EdS) or
  $3.5\times$ ($\Lambda$) (dotted line).  }
\label{fig:qsowf}
\end{figure}
\begin{figure} 
  \vspace{-2cm} \centerline{\hbox{\psfig{figure=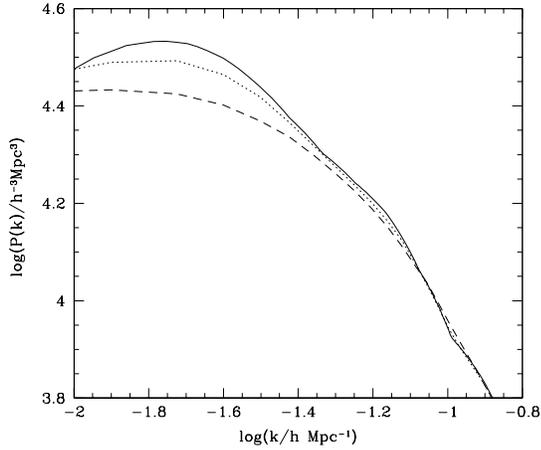,width=8.5cm}}}
\caption
{The effect of convolving a model power spectrum with the 2QZ $\Lambda$
  window function. The solid line shows a linear $\Lambda$CDM power
  specrum, with $\Omega_m=0.3$, $H_0=70$km\,s$^{-1}$Mpc$^{-1}$, and
  $\Omega_b=0.04$. The dotted line shows the same power spectrum
  convolved with the 2QZ $\Lambda$ window function, and the dashed line
  shows the power spectrum convolved with the 2dFGRS window function.
}
\label{fig:qsowf2}
\end{figure}

In Fig.~\ref{fig:qsowf} we show the window function, $|W_k|^2$, of the
2QZ, measured assuming either EdS or $\Lambda$. The window function is
anisotropic, due to the shape of the survey and variations in
completeness, described by the selection function.  In order to compare
model power spectra with that determined here for the 2QZ, we need to
convolve the model power spectra with the window function. In the case
where the power is isotropic, the same result would be obtained by
convolving with the spherically averaged window function. In reality,
whilst we expect the real-space power spectrum to be isotropic,
redshift-space distortions introduce small anisotropies into the shape
of the redshift-space 2QZ power spectrum (see Outram et al. 2001).
However, the {\it Hubble Volume} simulations suggest that these
anisotropies have little effect on the spherically averaged power
spectrum estimation, and so we only consider the spherically averaged
window function from here on. We also show an analytic fit to the
spherically averaged window function in Fig.~\ref{fig:qsowf}:

\begin{equation}
<|W_k|^2>_{EdS}=[1+(k/0.00204)^2+(k/0.00362)^4]^{-1}
\end{equation}
\begin{equation}
<|W_k|^2>_{\Lambda}=[1+(k/0.00132)^2+(k/0.00258)^4]^{-1}
\end{equation}

Due to the huge volume probed by 2QZ, the window function is much more
compact than that of galaxy surveys, and it takes the form of a steep
power law, proportional to $k^{-4}$ out to scales of $\approx 600
h^{-1}$Mpc ($\log (k / h \,{\rm Mpc}^{-1} ) \approx -2 $) for
$\Lambda$. For comparison we also show an analytic fit to the 2dFGRS
window function (Percival et al. 2001) in Fig.~\ref{fig:qsowf}. The 2QZ
window function has a very similar shape to that of the 2dFGRS, but
shifted to larger scales; a factor of $2.5\times$ larger for the 2QZ
EdS window function, or a factor of $3.5\times$ for $\Lambda$.

Fig.~\ref{fig:qsowf2} demonstrates the effect that the 2QZ window
function has on power spectra. On scales $< 200 h^{-1}$Mpc ($\log (k /
h \,{\rm Mpc}^{-1} ) > -1.5 $), the window function has very little
effect on the true power spectrum, whilst on scales larger than this
the true power is slightly suppressed. Even out to scales of $\approx
600 h^{-1}$Mpc ($\log (k / h \,{\rm Mpc}^{-1} ) \approx -2 $), the
effect of the window function is considerably smaller than the
statistical errors on the power spectrum estimate. The broader 2dFGRS
window function has a much larger effect, not only suppressing power on
scales $ > 100 h^{-1}$Mpc, but also smoothing out any baryonic
'wiggles' at smaller scales (Miller, Nichol \& Chen 2002), and
introducing covariance between the $P(k)$ data points.
 
\subsection{Error determination}
\label{sec:errors}

The errors on the power spectrum determination are estimated using the
method of Feldman, Kaiser \& Peacock (1994; FKP), equation 2.3.2. The
error is given by
\begin{equation} 
\frac{\sigma^2(k)}{P^2(k)} = \frac{(2\pi)^3 [1 +\frac{1}{n(r)P(k)}]^2}{V_k V_s}
\label{eq:fkper}
\end{equation}
where $V_k$ is the volume of each bin in $k$-space, estimated by $V_k=
N_k (\Delta k)^3 $ with $N_k$ the number of independent modes in the
$k$-shell and $(\Delta k)^3$ the volume of one $k$-mode. $V_s$ is the
volume of the survey. (We have assumed that the QSOs carry equal weight
when estimating the power spectrum and associated errors).  Hoyle
(2000) demonstrated that over a wide range of scales these errors were
found to be in reasonable agreement with errors estimated from the
dispersion over power spectrum estimates from the {\it Hubble Volume}
mock catalogues. On the smallest scales, non-linearities may lead to a
larger error than the FKP error estimate (Meiksin \& White 1999).

\section{The QSO power spectrum}
\label{sec:qsoreslts}
\begin{figure*}
\begin{minipage}{15cm}
  \vspace{-2cm}
  \centerline{\hbox{\psfig{figure=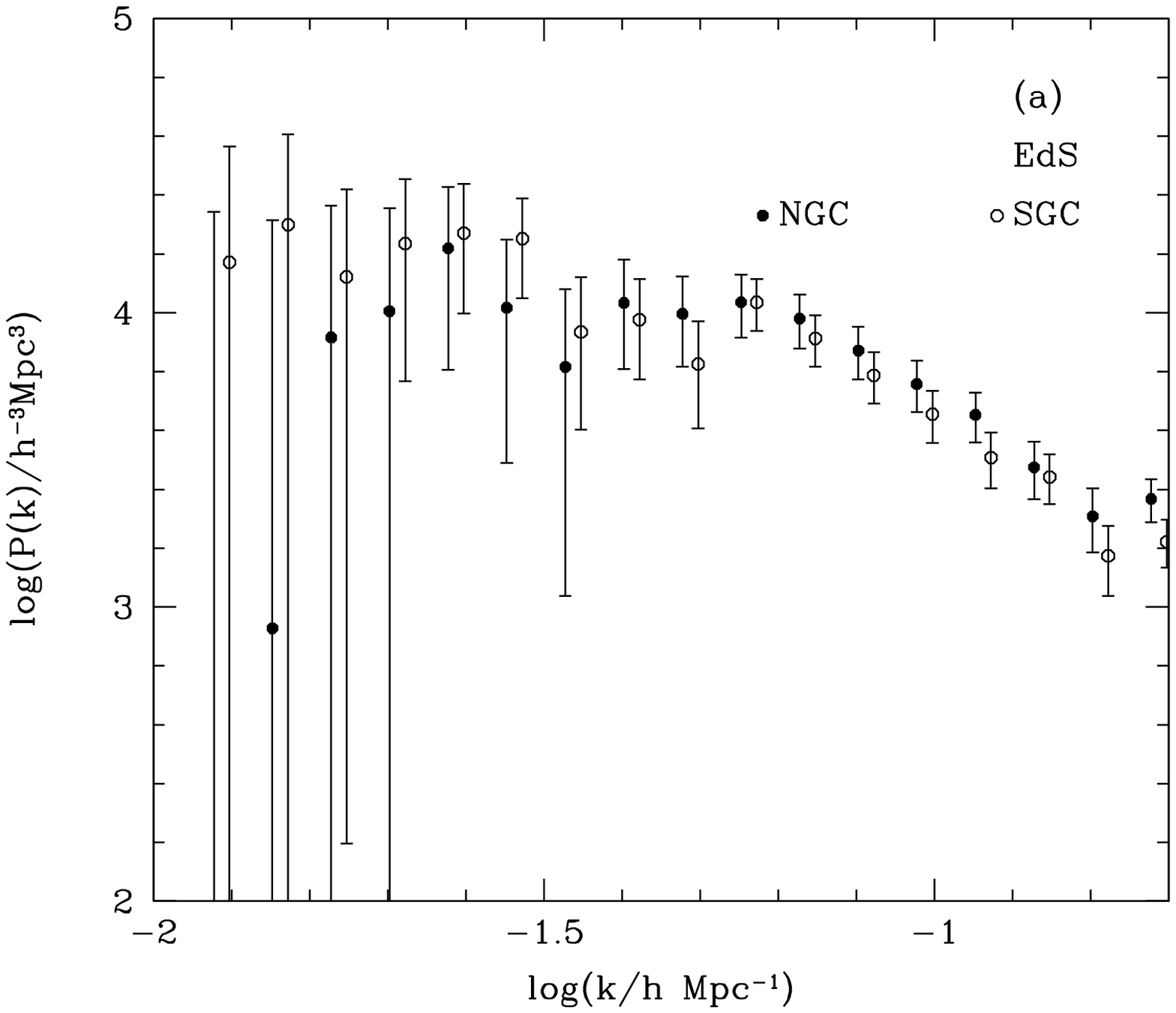,width=8.5cm}\psfig{figure=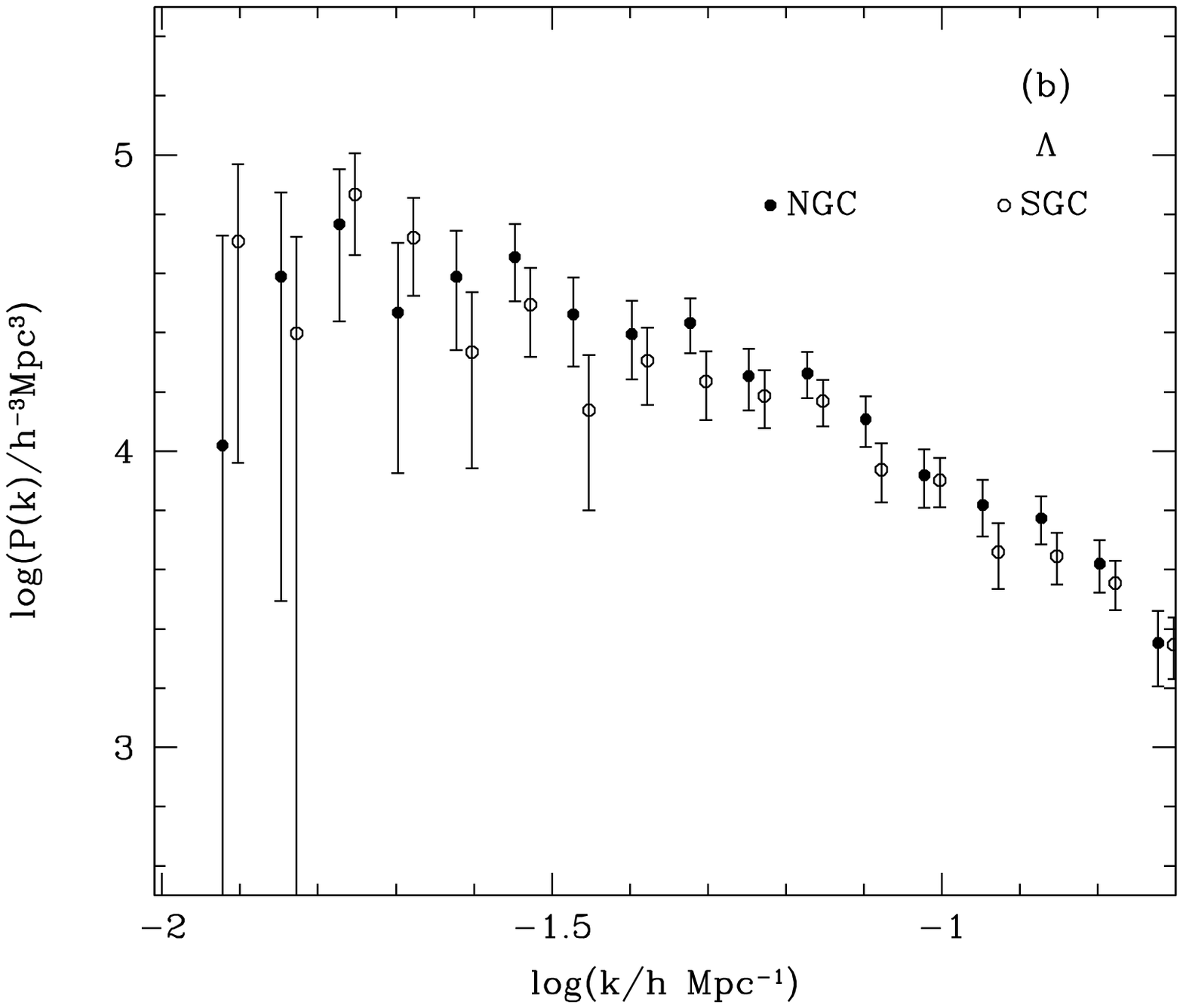,width=8.5cm}}}
  \vspace{-2cm}
  \centerline{\hbox{\psfig{figure=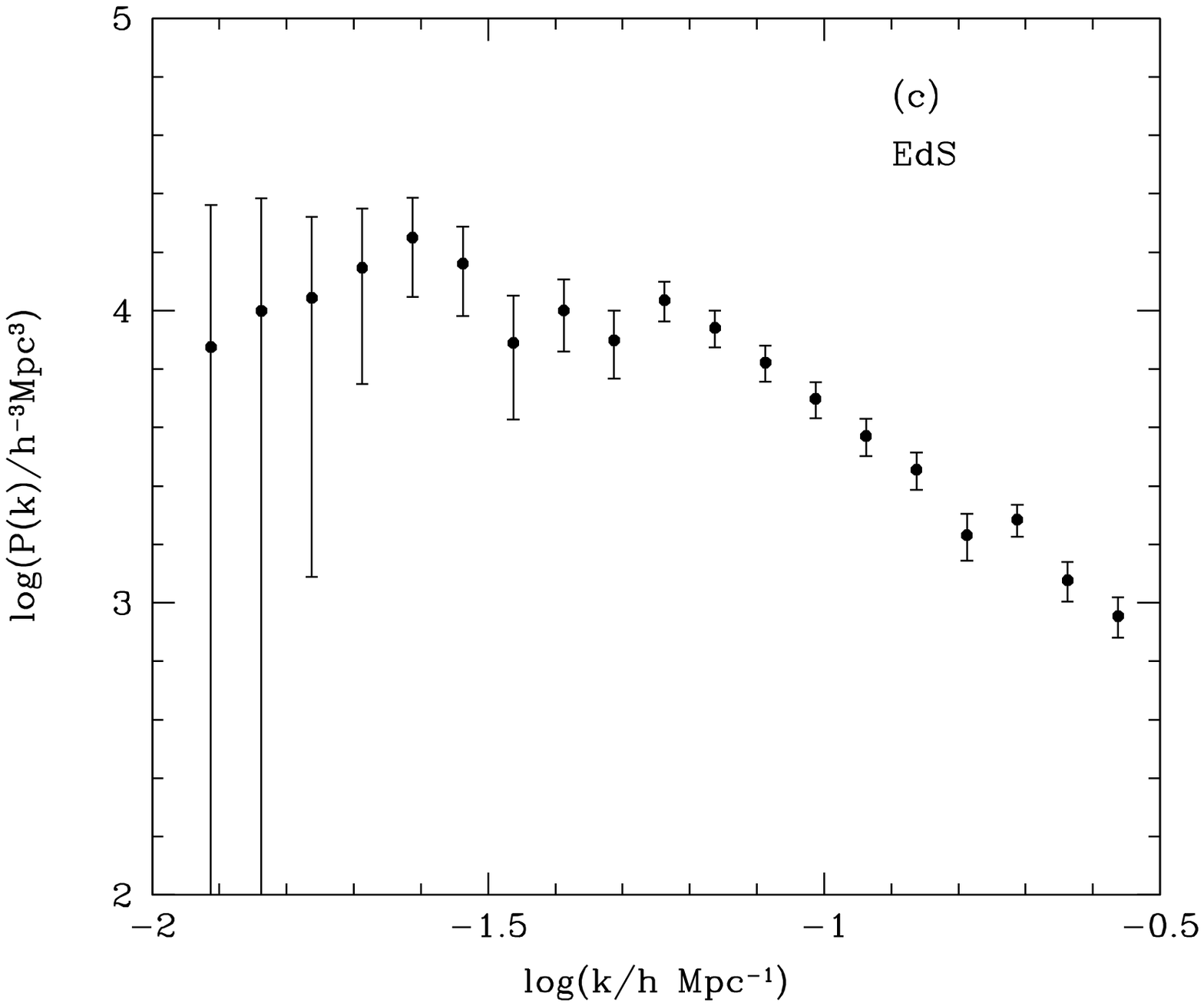,width=8.5cm}\psfig{figure=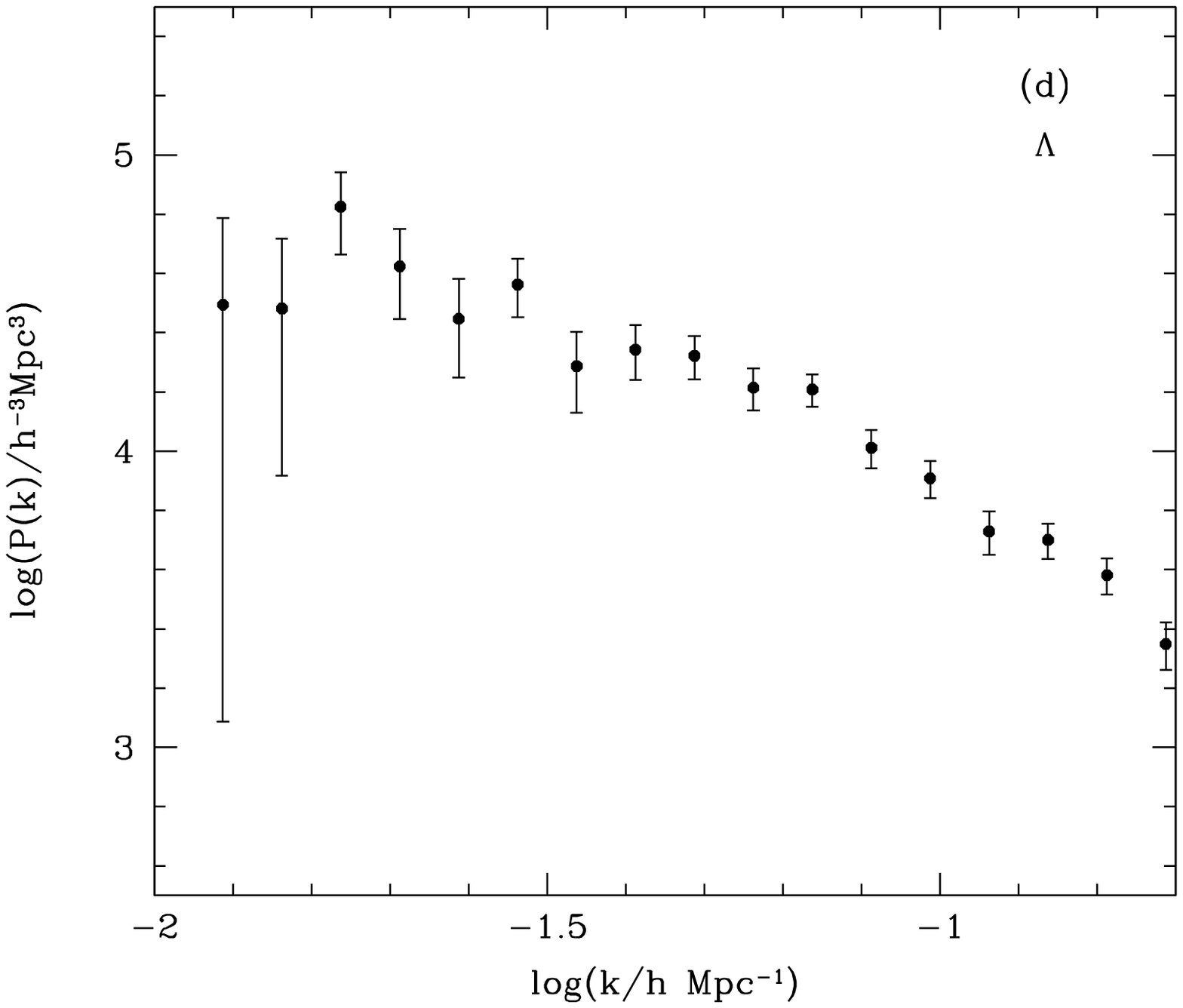,width=8.5cm}}}
\caption
{In Figures (a) and (b) we compare the power spectrum from the NGC
  (filled circles) and SGC (open circles).  Both sets of points have
  been offset by 0.01 in $\log (k/ h {\rm Mpc}^{-1})$ for clarity. In
  Figures (c) and (d) we show the combined QSO power spectra.  The EdS
  cosmology $r(z)$ is assumed for (a) and (c) and $\Lambda$ for (b) and
  (d).}
\label{fig:pk_ns}
\end{minipage}
\end{figure*}

In Figure \ref{fig:pk_ns} we show the power spectrum estimate obtained,
using the method outlined above, for the 2QZ dataset restricted to QSOs
with redshift in the range $0.3<z<2.2$. We compare the NGC and SGC
power spectra measured assuming the EdS (a) or $\Lambda$ (b).  The
combined power spectra are also shown, assuming EdS (c) or $\Lambda$
(d). The power spectra measured from the two strips agree well over a
wide range of scales, with a dispersion that is consistent with our
error estimate. The errors shown here and throughout the paper are
1$\sigma$ errors estimated using the FKP method. The power spectrum
estimates obtained are also given in Table~\ref{table2} (assuming EdS)
and Table~\ref{table3} (assuming $\Lambda$).

\begin{figure} 
  \vspace{-2cm} \centerline{\hbox{\psfig{figure=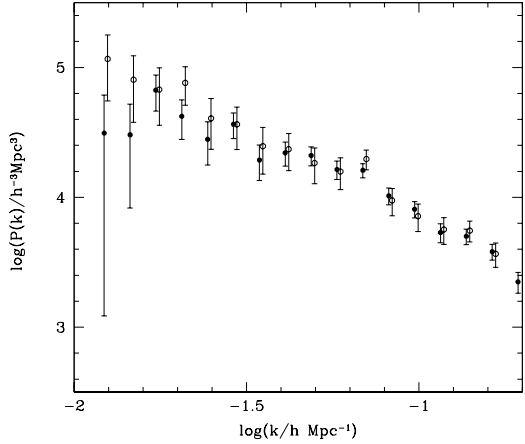,width=8.5cm}}}
\caption
{A comparison between the 2QZ power spectrum (assuming $\Lambda$) from
  the final sample (filled circles) with that derived from the initial
  10k catalogue (open circles; Hoyle et al. 2002).  Both sets of points
  have been offset by 0.01 in $\log (k/ h {\rm Mpc}^{-1})$ for clarity.
  There is very good agreement, except on the largest scales where the
  excess power was introduced by the complicated incomplete window
  function in the 10k sample.  }
\label{fig:pk10k}
\end{figure}

We compare this power spectrum estimate (assuming $\Lambda$) to that
obtained by Hoyle et al. (2002) using the 10k 2QZ catalogue in
Fig.~\ref{fig:pk10k}. On scales smaller than $\log (k / h \,{\rm
  Mpc}^{-1})\sim -1.6$, the spectra are very similar, with the main
difference being the reduced errors for the new estimate due to the
increased sample size. On large scales, however, the 10k power spectrum
estimate is slightly higher than the latest determination. This
over-estimate was largely due to problems accounting for the
complicated window of the incomplete 10k survey. However, with this
difference in mind we now examine some of the other potential
contributions to systematic errors in the power spectrum determination
on large scales.

\begin{table}
\caption{Values of the QSO power spectrum $P(k)$ estimated from QSOs
  with redshift in the range $0.3<z<2.2$ contained in the 2dF QSO
  Redshift Survey Catalogue, assuming an EdS cosmology $r(z)$. The
  errors are estimated using the method of Feldman, Kaiser \& Peacock (1994).  }
\begin{center}
\begin{tabular}{@{}ccc}
\hline
$\log (k / h \,{\rm Mpc}^{-1} )$&$P(k) / h^{-3} \,{\rm Mpc}^{3}$&$\Delta P(k) / h^{-3} \,{\rm Mpc}^{3} $\\
\hline
    -1.91250 &   7497.86 &   15528.9\\
    -1.83750 &   9975.61 &   14241.7\\
    -1.76250 &   11055.6 &   9826.21\\
    -1.68750 &   14009.9 &   8392.96\\
    -1.61250 &   17765.4 &   6605.29\\
    -1.53750 &   14482.3 &   4921.19\\
    -1.46250 &   7753.39 &   3518.10\\
    -1.38750 &  10000.78 &   2758.25\\
    -1.31250 &   7918.47 &   2079.62\\
    -1.23750 &   10850.0 &   1668.21\\
    -1.16250 &   8722.57 &   1252.32\\
    -1.08750 &   6644.70 &   940.720\\
    -1.01250 &   4991.80 &   710.215\\
   -0.937500 &   3716.06 &   538.652\\
   -0.862500 &   2851.89 &   411.061\\
   -0.787500 &   1703.66 &   312.109\\
   -0.712500 &   1924.72 &   241.619\\
   -0.637500 &   1194.02 &   184.568\\
   -0.562500 &   900.635 &   141.926\\
\hline
\end{tabular}
\end{center}

\label{table2}
\end{table}

\begin{table}
\caption{Values of the QSO power spectrum $P(k)$ estimated from QSOs
  with redshift in the range $0.3<z<2.2$ contained in the 2dF QSO
  Redshift Survey Catalogue, assuming a $\Lambda$ cosmology $r(z)$. The
  errors are estimated using the method of Feldman, Kaiser \& Peacock (1994).  }
\begin{center}
\begin{tabular}{@{}ccc}
\hline
$\log (k / h \,{\rm Mpc}^{-1} )$&$P(k) / h^{-3} \,{\rm Mpc}^{3}$&$\Delta P(k) / h^{-3} \,{\rm Mpc}^{3} $\\
\hline
    -1.98750 &    34308.1  &   40765.3\\
    -1.91250 &    31228.7  &   30009.9\\
    -1.83750 &    30285.0  &   22002.5\\
    -1.76250  &   66904.5  &   20668.8\\
    -1.68750 &    42073.0 &    14148.3\\
    -1.61250 &    27967.6 &   10203.46\\
    -1.53750 &    36571.7  &   8169.82\\
    -1.46250 &    19371.4 &    5884.91\\
    -1.38750 &    22035.7 &    4602.21\\
    -1.31250 &    20998.8 &    3522.13\\
    -1.23750 &   16388.0 &   2646.94\\
    -1.16250 &   16152.2 &   2040.24\\
    -1.08750 &  10279.27 &   1531.78\\
    -1.01250 &   8100.51 &   1171.00\\
   -0.937500 &   5352.25 &   891.509\\
   -0.862500 &   5013.59 &   687.006\\
   -0.787500 &   3816.35 &   527.254\\
   -0.712500&    2233.81 &   403.834\\
   -0.637500 &   841.037 &   309.547\\
   -0.562500 &   800.378 &   238.849\\
\hline
\end{tabular}
\end{center}

\label{table3}
\end{table}

\subsection{Large-scale systematic errors}

\label{pksys}

\begin{figure} 
  \vspace{-2cm} \centerline{\hbox{\psfig{figure=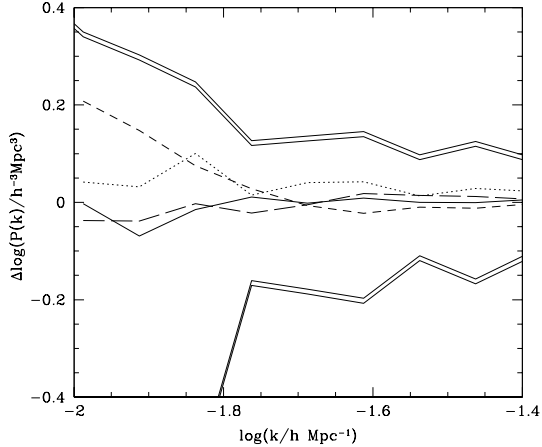,width=8.5cm}}}
\caption
{The effect of making several changes to the observation mask are
  demonstrated. The solid line shows the change in the estimated power
  spectrum when SDSS photometry is used to calibrate the UKST plates in
  the NGP, rather than our own calibration. The short dashed line shows
  the effect of not correcting for dust. The long dashed line shows the
  difference between the power spectrum determinations using two
  different spectroscopic completeness estimates. The dotted line shows
  the effect of including a spectroscopic completeness estimate when
  considering only a sample with spectroscopic completeness $>85$ per
  cent. For comparison, the double lines show the 1$\sigma$ errors for
  the power spectrum estimation.  }
\label{fig:pksys}
\end{figure}

One possible source of systematic errors is the uncertainty in the
zero-point calibration of the UKST field plates that were used to
define the input catalogue for the 2QZ survey. The fields are
$5\deg\times5\deg$, and so any differences in the plate calibration
could introduce power on the largest scales probed. The recent release
of the SDSS EDR data (Stoughton et al. 2002) has given us the
opportunity to apply a new calibration to our data in the NGP to test
this effect, after converting the Sloan colours to our bands (Blanton
et al. 2001; Fukugita, Shimasaku, \& Ichikawa 1995). A comparison of
the two calibrations suggests that our calibration introduces
plate-to-plate uncertainty in the QSO number density of the order of 4
per cent. In Fig.~\ref{fig:pksys} we show the effect of using the SDSS
calibration in the NGP on our power spectrum estimate. The difference
in $\log P(k)$ determined using the two calibrations is shown by the
solid line. On scales $\log (k / h \,{\rm Mpc}^{-1})> -1.8$ there is
virtually no difference. On large scales, using the SDSS calibration
would lead to a small decrease in the measured power spectrum.
Unfortunately we have no similar data to re-calibrate the SGP strip,
and so, because this effect is very small, we have continued to use our
original calibration.

A second potential source of systematic error is Galactic dust. We
account for dust extinction using the estimates given by Schlegel,
Finkbeiner \& Davis (1998). If dust extinction is not correctly
accounted for then this could introduce large scale variations in the
observed QSO number density, and hence power.  The short-dashed line in
Fig.~\ref{fig:pksys} shows a comparison between the power spectrum
estimates with and without a dust correction. Applying the correction
significantly reduces power on scales $\log (k / h \,{\rm Mpc}^{-1})<
-1.8$. Whilst this gives us confidence in the dust maps, their
resolution is fairly low, and so dust could still affect our estimate
on smaller scales.

Finally we consider the effect of different choices of spectroscopic
incompleteness corrections. In the 10k power spectrum analysis, only
regions with spectroscopic completeness in excess of 85 per cent were
considered, and no further correction was applied to account for any
QSOs still missing in the unidentified spectra. As identification is
dependent on the quality of the spectra, which varies with each
pointing due to observing conditions, the fraction of missing QSOs
could vary slightly from region to region even in this 85 per cent
complete sample, introducing large scale power. To investigate this we
compare the power spectra determined using an 85 per cent spectroscopic
completeness cut, then a further correction for observational
completeness (as carried out in the 10k analysis) to one with the same
85 per cent cut, but also correcting for spectroscopic completeness,
using the same form of correction as in this paper; assuming that
fraction of QSOs in unidentified objects is the same as the fraction in
the identified spectra. The result is shown by the dotted line in
Fig.~\ref{fig:pksys}. The lack of a spectroscopic completeness
correction introduces a slight over-estimate in the power, especially
at large scales. This may also partly explain the difference between
the 10k and final power spectrum estimates.

The spectroscopic completeness correction we apply simply assumes that
the spectroscopic incompleteness of the QSOs is the same as that of the
full input catalogue, and hence that $\sim59$ per cent of unidentified
objects are QSOs. Whilst this is favoured by evidence from
re-observations, there is still uncertainty in the proportion of QSOs
remaining in the unidentified spectra, and a slightly lower fraction of
unidentified QSOs is still compatible with the observations. To test
the effect of this uncertainty in the spectroscopic completeness
correction we can construct a new completeness map where we instead
assume that a lower fraction of approximately 40 per cent of objects
that remain unidentified after observation are QSOs. A comparison
between the power estimated using the two completeness corrections is
shown by the long-dashed line in Fig.~\ref{fig:pksys}. The difference
in power is very small on all scales, indicating that the uncertainty
in the unidentified QSO fraction has a negligible effect on the power
spectrum estimate.

It is also possible that there are spectroscopic completeness
variations across individual 2dF fields, which could in principle
introduce power on smaller scales. By artificially introducing
completeness gradients across the 2dF fields in the random catalogue,
we can investigate the likely size of any systematic uncertainty due to
this effect. Even with a completeness gradient of 20 per cent from the
centre to the edge of each field, the change in the measured power
spectrum is no larger than that observed for any of the other potential
systematics discussed above. As this artificially introduced gradient
is much larger than the expected size of completeness variations across
2dF fields, given that the overall spectroscopic completeness for the
survey is 86 per cent (and this is largely due to other effects such as
weather conditions and seeing), we conclude that the power added by
such an effect would be small, and we do not attempt to correct for
this.

All of the potential systematics discussed above are much smaller than
the statistical errors on all scales. With the exception of the dust
correction, they are of the order of $\Delta\log P(k) \sim 0.05$ even
on scales of $\log (k / h \,{\rm Mpc}^{-1}) \sim -2$. Therefore we have
confidence that our estimate is robust out to these scales.

\subsection{The shape of the QSO power spectrum}
\label{sec:gam}

\begin{figure} 
  \vspace{-2cm} \centerline{\hbox{\psfig{figure=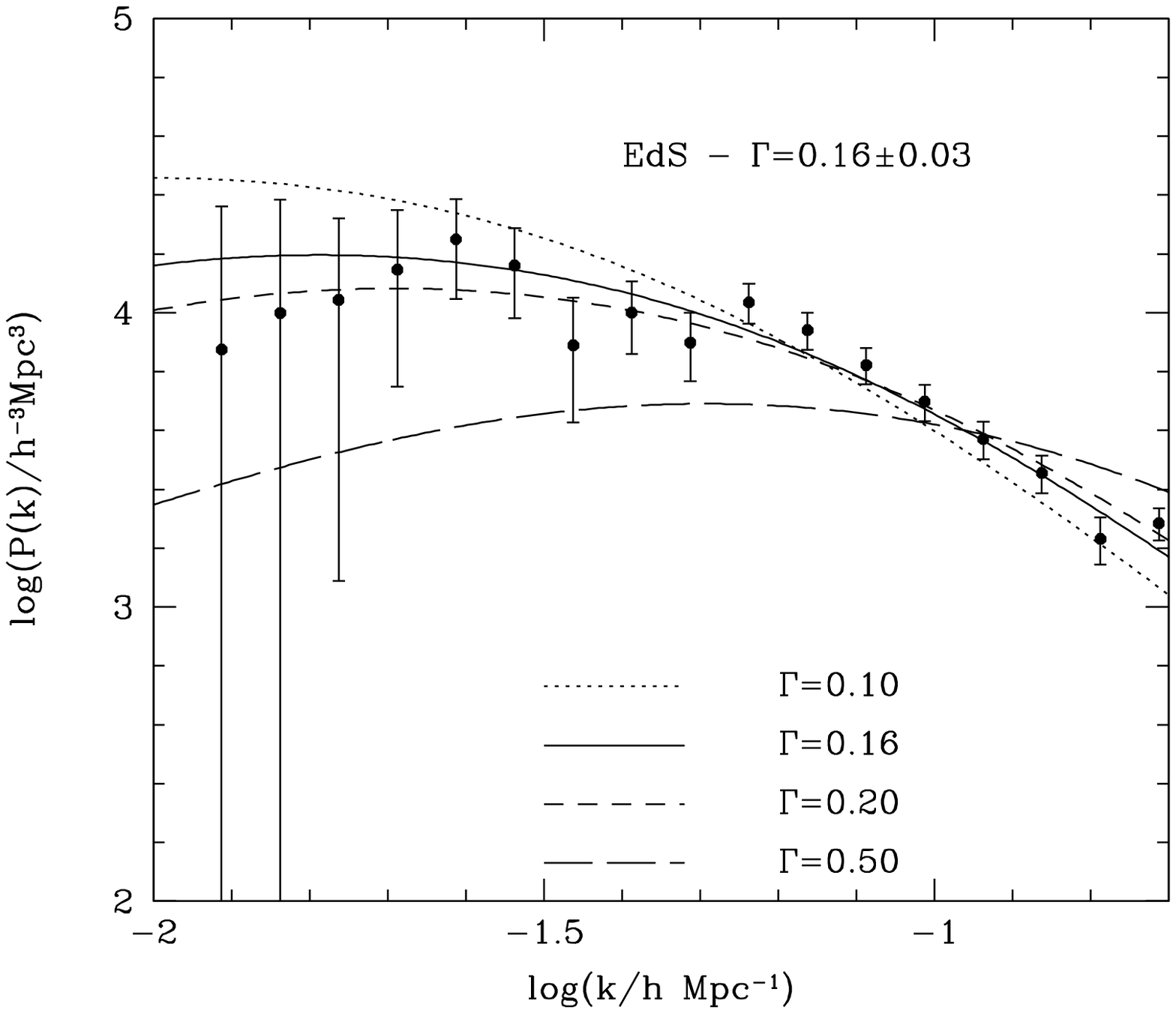,width=8.5cm}}}
  \vspace{-2cm} \centerline{\hbox{\psfig{figure=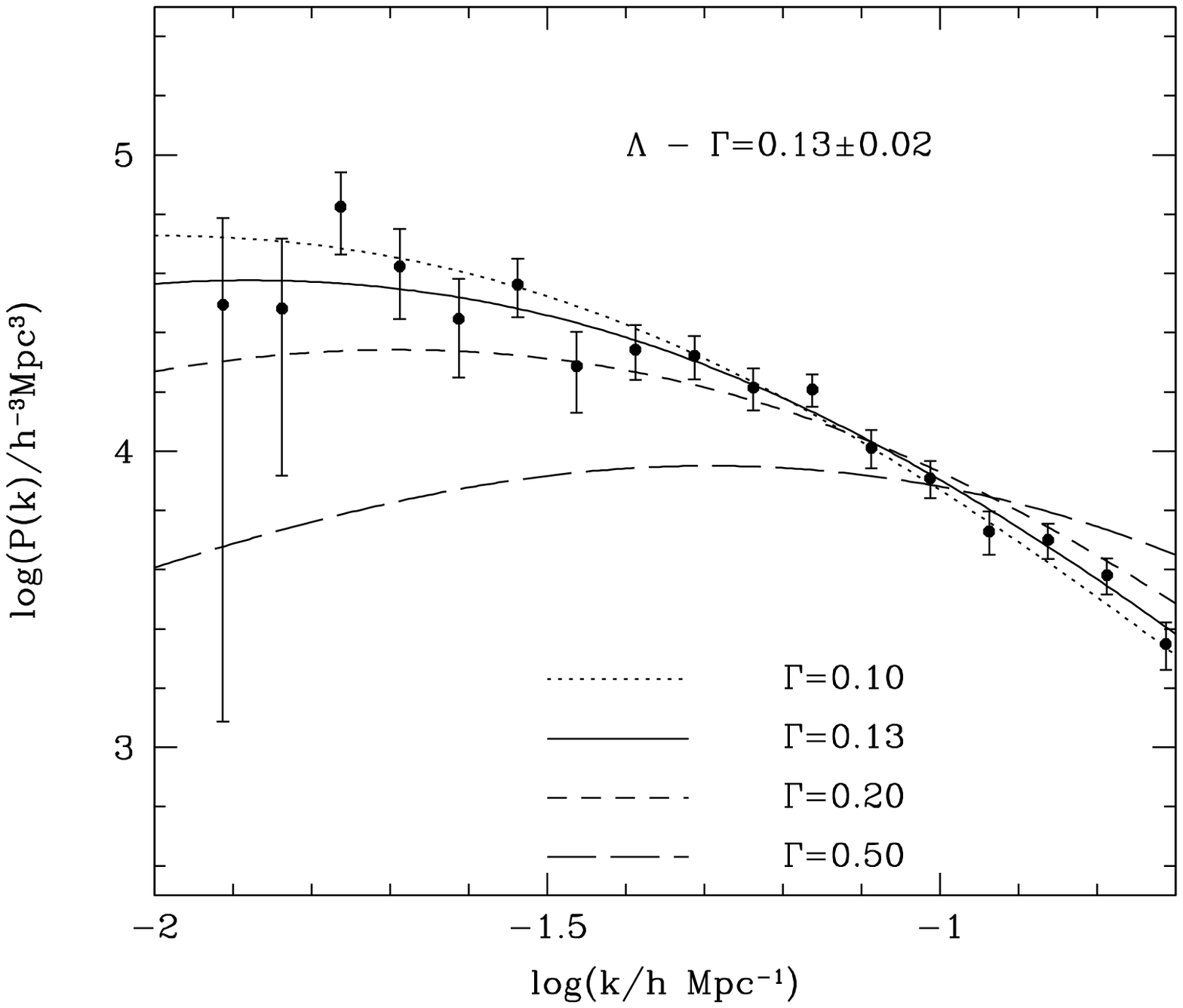,width=8.5cm}}}
\caption
{ Model power spectra with varying shape parameter, $\Gamma$, are
  compared to the QSO power spectra for the two choices of cosmology.
  In the upper plot, the solid points show the QSO power spectra with
  the EdS $r(z)$. Overlaid is the best-fitting $\Gamma$ model, with
  $\Gamma=0.16$ (solid line). In the lower plot, the solid points show
  the QSO power spectra assuming the $\Lambda$ $r(z)$. Overlaid is the
  best-fitting $\Gamma$ model, with $\Gamma=0.13$ (solid line). On both
  plots, the dotted line shows a $\Gamma=0.1$ model, the short dashed
  line shows a $\Gamma=0.2$ model, and the long dashed line shows a
  $\Gamma=0.5$ model. The latter model is favoured by a Standard CDM
  model, but clearly rejected by the data.  }
\label{fig:pkgam}
\end{figure}

In CDM models, the power spectrum should turn over at a scale which
depends on the shape parameter, $\Gamma$, of the power spectrum. The
shape of the power spectrum and the position of the turnover contains
information on cosmological parameters that can lead to strong
constraints. In pure CDM models (i.e. without baryons) with a
scale-invariant spectrum of initial fluctuations $\Gamma=\Omega_{\rm
  m}h$. However, the inclusion of baryons, or a tilted spectrum of
initial fluctuations complicates the picture, so a measurement of
$\Gamma$ alone cannot be inverted reliably to give $\Omega_{\rm m}h$
(Sugiyama 1995, Peacock 2000). For example, the $\Lambda$ model we have
assumed ($\Omega_{\rm m}$=0.3, $\Omega_{\Lambda}$=0.7) to calculate the
$r(z)$ would have a value of $\Gamma=0.21$, if we further assume that
$H_0\sim70$km\,s$^{-1}$Mpc$^{-1}$ (e.g. Tanvir et al. 1995; Mould et
al. 2000), in the absence of baryons. However, when baryons are taken
into account this model would then have an effective value of
$\Gamma=0.17$. For a full comparison, we also need to consider the
effects of the window function, and redshift-space distortions. We have
had to estimate the real-space distances to the QSOs from their
observed redshift, and peculiar velocities, large-scale flows,
geometrical assumptions and uncertainties in redshift measurements can
all affect the shape and amplitude of the power spectrum we have
determined. Despite this, $\Gamma$ is a very useful parameter for
describing the general shape of power spectra, and allowing a simple
comparison between different measurements or models.

To parametrize the QSO power spectra we have measured, we fit model CDM
power spectra with different values of the shape parameter $\Gamma$. We
assume the fit to the CDM transfer function given by Bardeen et al.
(1986). For each value of $\Gamma$, we allow the amplitude to vary
freely, and minimize the $\chi^2$ of the fit. The best fit models are
shown in Fig.~\ref{fig:pkgam}. For an EdS $r(z)$, we find that models
with $\Gamma=0.16\pm0.03$ (1$\sigma$ errors) provide the best fit to
the power spectrum, whereas assuming a $\Lambda$ $r(z)$, the power
spectrum is slightly steeper, with $\Gamma=0.13\pm0.02$. These low
measurements of $\Gamma$ indicate significant large-scale power,
considerably in excess of Standard ($\Omega_{\rm m}=1$) CDM model
predictions, and consistent with the $\Lambda$ model prediction
discussed above. In the case of the $\Lambda$ $r(z)$, whilst the power
spectrum flattens at the largest scales probed, we see no evidence of
the expected turnover below scales of $\sim$400$h^{-1}$Mpc ($\log
k\sim-1.8$). However, in the slightly flatter EdS $r(z)$ power spectrum
there is a hint of a possible turnover by scales of
$\sim$400$h^{-1}$Mpc. In either case we can strongly rule out a
turnover in the power spectrum at scales below $\sim$300$h^{-1}$Mpc
($\log k\sim-1.7$). We will consider the effect of baryons, the window
function, and redshift-space distortions in Section~\ref{sec:qsoimp} in
order to derive constraints on $\Omega_{\rm m}h$ from the shape of the
power spectrum.

\subsection{Comparison with P(k) from galaxy and cluster surveys}
\label{comp}

\begin{figure}
  \vspace{-2cm} \centerline{\hbox{\psfig{figure=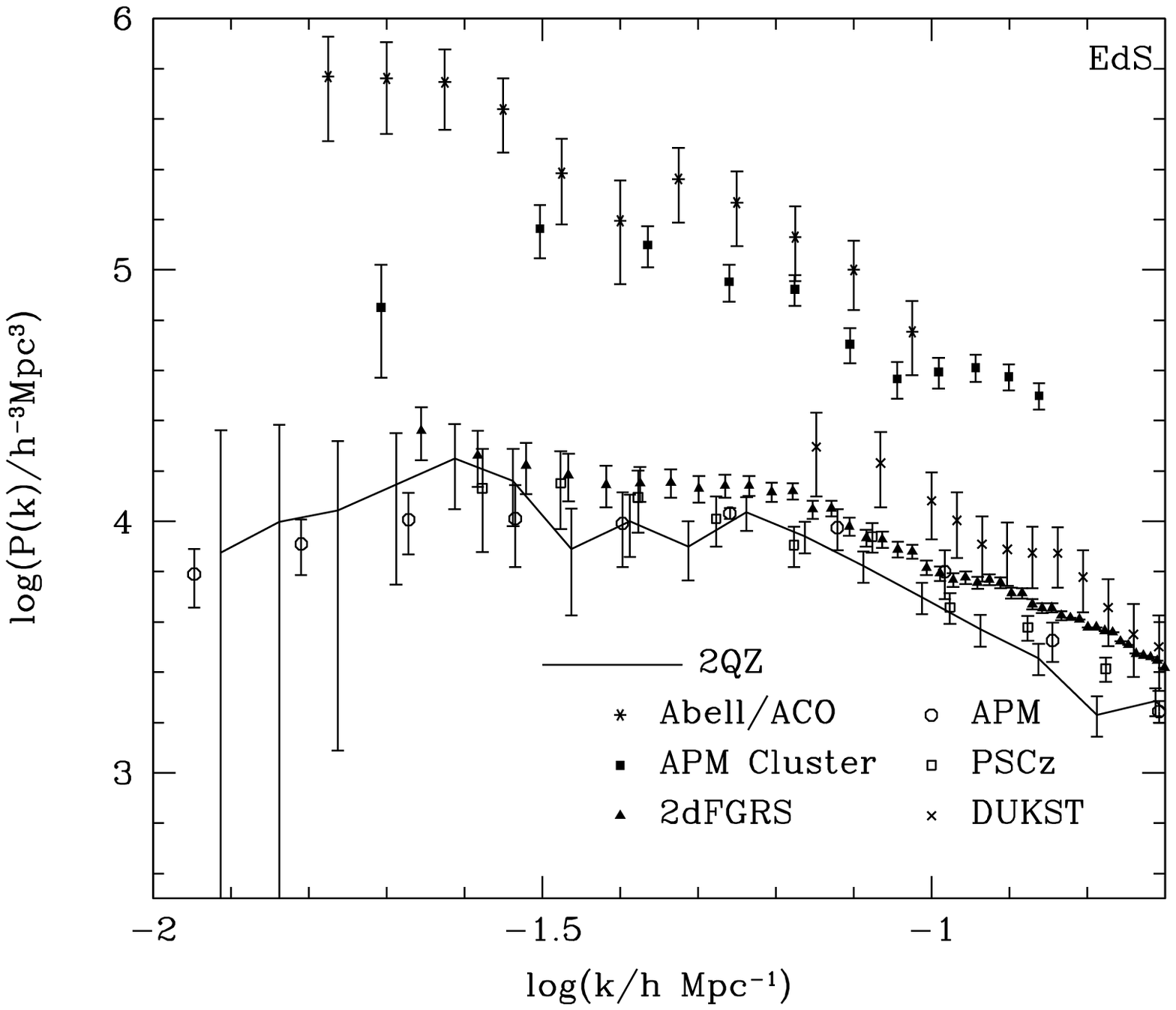,width=8.5cm}}}
  \vspace{-2cm} \centerline{\hbox{\psfig{figure=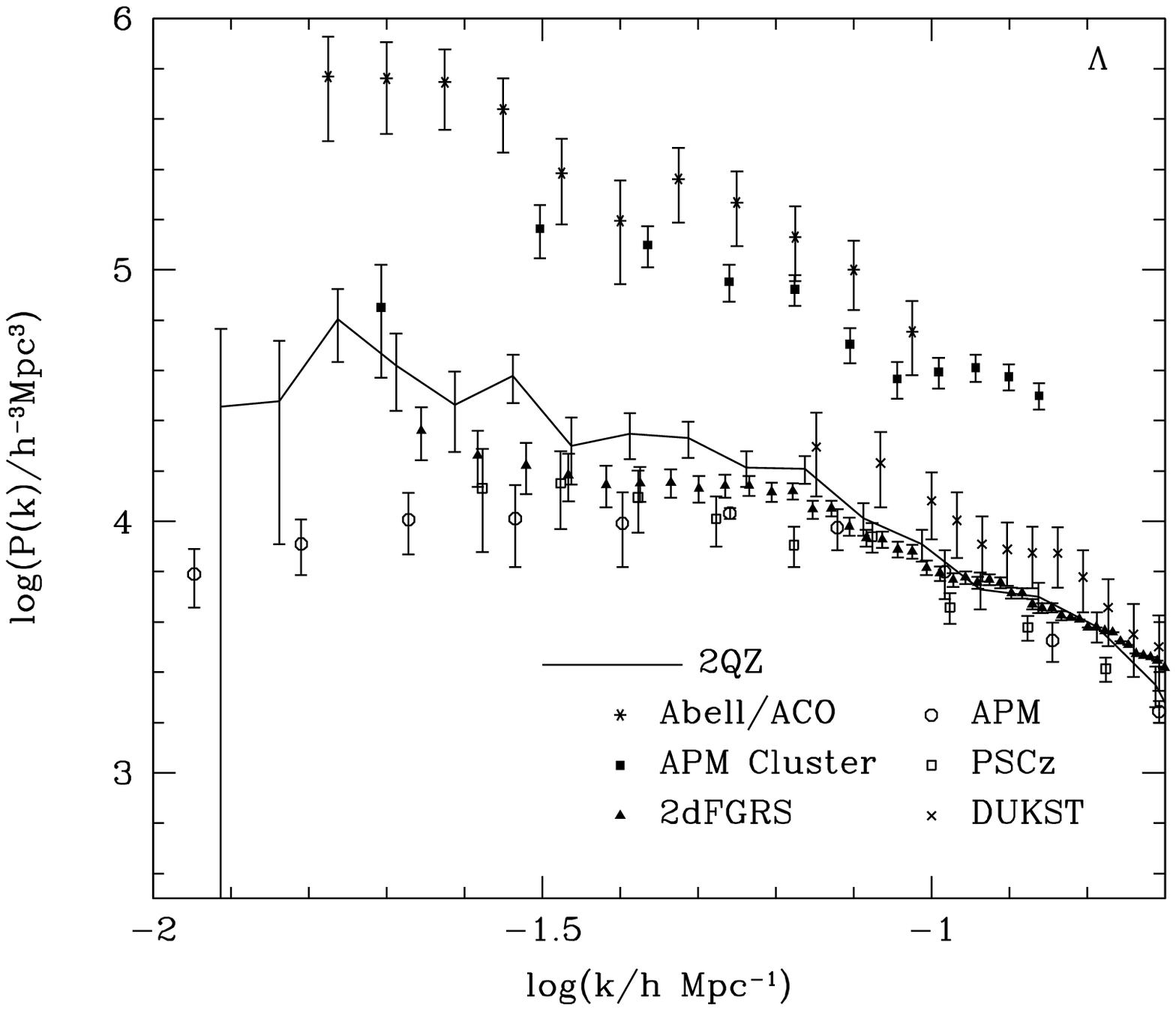,width=8.5cm}}}
\caption
{ A comparison of galaxy and QSO power spectra. The EdS $r(z)$ is
  assumed in the upper plot and the $\Lambda$ $r(z)$ is assumed in the
  lower plot. The solid line shows the QSO power spectra.  The errors
  are 1$\sigma$ FKP errors. The points show the power spectra of local
  galaxies, determined from the 2dFGRS (Percival et al. 2001), PSCz
  (Sutherland et al. 1999), APM (Gaztanaga \& Baugh 1998), and DUKST
  (Hoyle et al. 1999) galaxy surveys and the power spectra of local
  Abell/ACO clusters (Miller \& Batuski 2001) and APM clusters (Tadros,
  Efstathiou \& Dalton 1998)}
\label{fig:pscomp}
\end{figure}

We compare the QSO power spectrum with the power spectra of local
galaxies, determined from the 2dFGRS (Percival et al. 2001), PSCz
(Sutherland et al. 1999), APM (Gaztanaga \& Baugh 1998), and DUKST
(Hoyle et al. 1999) galaxy surveys and the power specta of local
Abell/ACO clusters (Miller \& Batuski 2001) and APM clusters (Tadros,
Efstathiou \& Dalton 1998) in Fig.~\ref{fig:pscomp}. Assuming a
$\Lambda$ $r(z)$, the amplitude of QSO clustering at $\bar{z}\sim1.4$
is similar to that of local galaxies, and almost an order of magnitude
below that of galaxy clusters. Assuming an EdS $r(z)$, the amplitude of
QSO clustering is slightly lower than the local galaxy clustering
amplitude.  The $\Lambda$ $r(z)$ QSO power spectrum appears steeper
than that of the 2dFGRS, with more power on large scales. This is
largely due, however, to the effects of the window function on the
2dFGRS power spectrum, as demonstrated in Fig.~\ref{fig:qsowf2}. For a
true comparison of the relative shapes of the galaxy and QSO power
spectra, we will fit models convolved with their respective window
functions in Section~\ref{sec:qsoimp}.

\subsection{The evolution of QSO clustering}
\label{sec:pkevo}

\begin{figure} 
  \vspace{-2cm} \centerline{\hbox{\psfig{figure=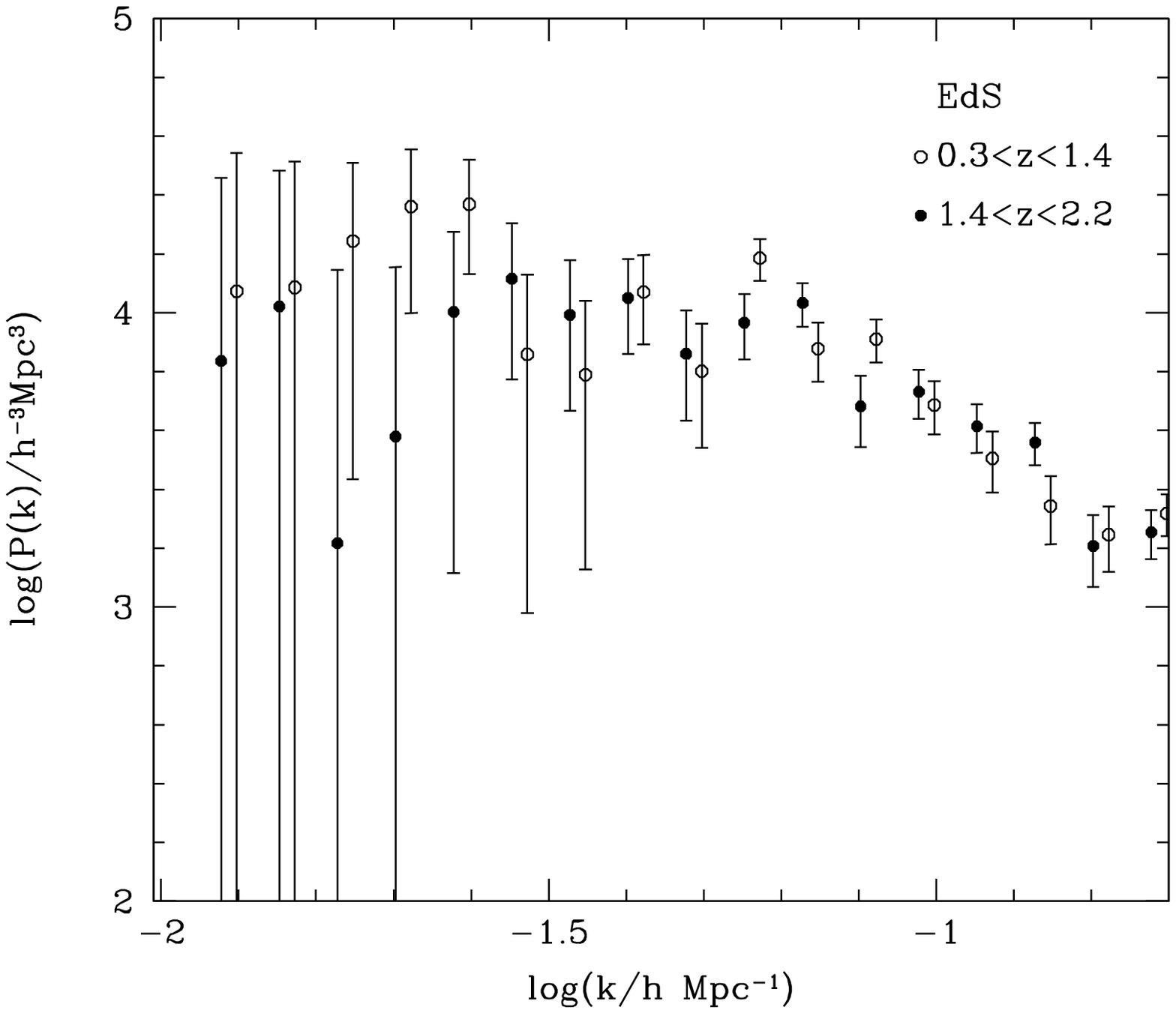,width=8.5cm}}}
  \vspace{-2cm} \centerline{\hbox{\psfig{figure=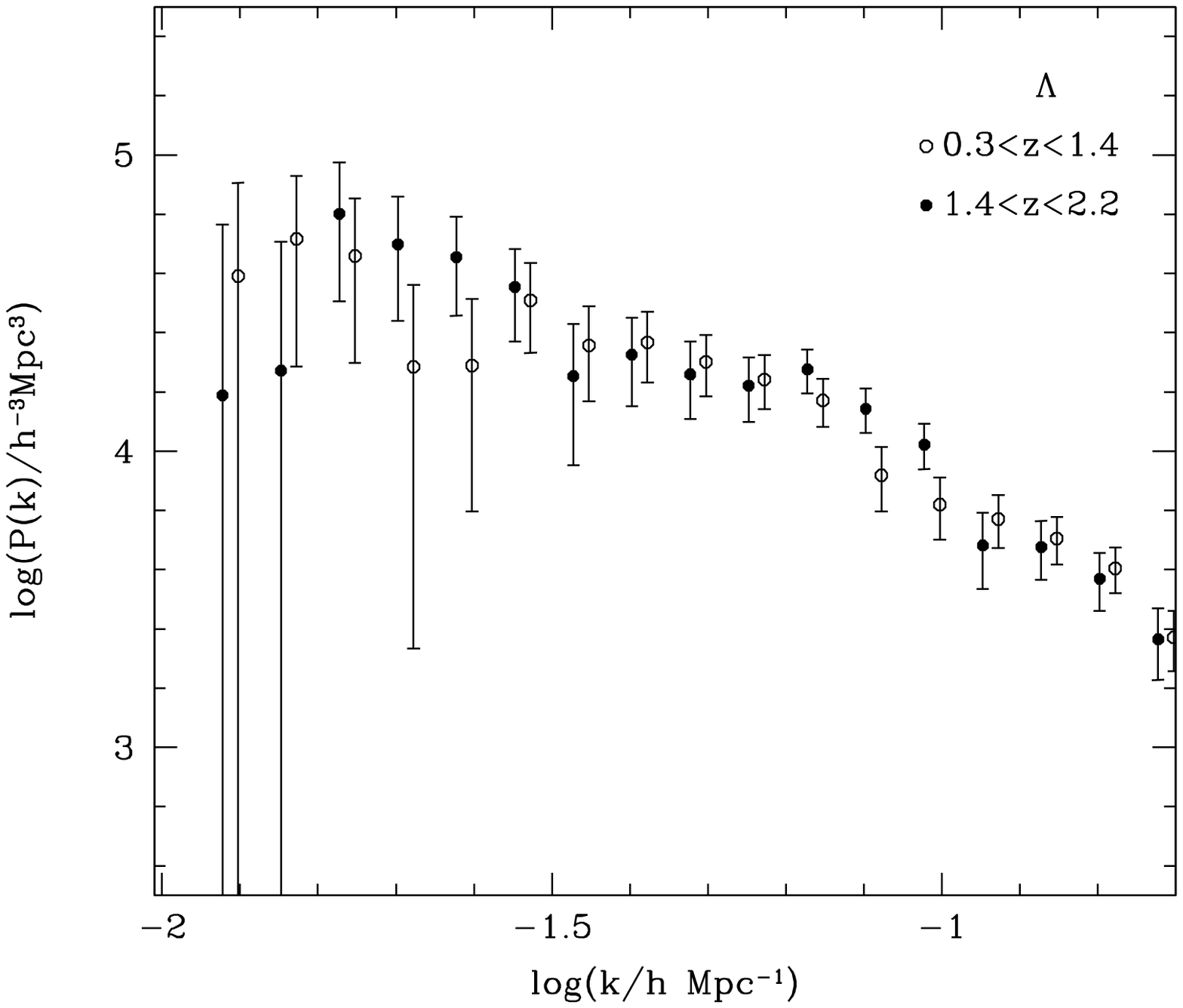,width=8.5cm}}}
\caption
{ The power spectrum of QSOs measured at different redshifts. The EdS
  $r(z)$ is assumed in the upper plot and $\Lambda$ is assumed in the
  lower plot. The open circles show the power spectrum of QSOs with
  redshifts in the range $0.3<z<1.4$ and the filled circles show the
  power spectrum of QSOs with redshifts in the range $1.4<z<2.2$. The
  errors are 1$\sigma$ FKP errors. The points are slightly offset for
  clarity.  }
\label{fig:pkevo}
\end{figure}

To investigate the effects of evolution on the amplitude of QSO
clustering we split the QSO sample into two, a high redshift sample
with $z>1.4$, and a low redshift sample with $z<1.4$, and compare their
respective power spectra. The results are shown in
Fig.~\ref{fig:pkevo}. Assuming either EdS or $\Lambda$, we see
virtually no evolution in the amplitude of clustering between the two
samples, at $0.3<z<1.4$ and $1.4<z<2.2$. To quantify this, we fit the
amplitude of the best fitting model power spectra for each cosmology
found in section~\ref{sec:gam} ($\Gamma=0.16$ for EdS and $\Gamma=0.13$
for $\Lambda$) separately for each redshift bin. The results are shown
in Table~\ref{table}.  We find that the amplitude of QSO clustering is
almost identical in the two redshift bins. Assuming EdS, the difference
in amplitude between the two redshifts is insignificant. Assuming
$\Lambda$, we marginally detect a slight fall in amplitude of $\Delta
\log P =0.05$ from high redshift to low redshift, at about 1$\sigma$
significance. This is in very good agreement with the results of the
2QZ correlation function analysis of Croom et al. (2001b).

We fixed the value of $\Gamma$ for the above analysis, as the shape
parameter is assumed not to evolve with redshift (in the linear
regime), in order to investigate the evolution of the amplitude of QSO
clustering. We can test this assumption by comparing the value of
$\Gamma$ in the low and high redshift bins. For the $\Lambda$ $r(z)$,
we find that this is indeed the case, with best fit values of
$\Gamma=0.14\pm0.03$ at $0.3<z<1.4$ and $\Gamma=0.13\pm0.03$ at
$1.4<z<2.2$.  For the EdS $r(z)$, however, there is marginal evidence
for a slight flattening of the power spectrum at higher redshift, with
best fit values of $\Gamma=0.15\pm0.05$ at $0.3<z<1.4$ and
$\Gamma=0.20\pm0.05$ at $1.4<z<2.2$. One possible, though unlikely
explanation for this observed flattening could be evolution of a
scale-dependent QSO bias on $\ga100$\,h$^{-1}$Mpc scales. If this were
the case it could be a major problem for the cosmological parameter
analysis in the next section, and other similar analyses that use
galaxies or QSOs as tracers of large-scale structure. Alternatively,
the flattening could be due to the different shape of the window
function in the two redshift bins. However, the two window functions
are very similar, and if anything, the high redshift window function is
slightly more compact, so we would expect less suppression of
large-scale power, not more, at high redshift.

If our assumption of an EdS cosmology $r(z)$ were incorrect, as
suggested by the low measured values of $\Gamma$ and hence $\Omega_mh$
(see Section~\ref{sec:qsoimp}) then we would have a simple explanation
for the observed flattening. If, for example, the $\Lambda$ cosmology
were the true cosmology, then our assumption of an EdS $r(z)$ would
lead to an underestimate of the comoving distance scale, by about 25
per cent in the low redshift bin, and 50 per cent in the high redshift
bin. If we apply this correction to distance scale of the power
spectrum determined assuming EdS, we then obtain consistent values of
$\Gamma=0.12\pm0.03$ at $0.3<z<1.4$ and $\Gamma=0.13\pm0.03$ at
$1.4<z<2.2$. The observed evolution in shape assuming an EdS $r(z)$
could therefore be taken as further marginal evidence against an EdS
cosmology.

\begin{table}
\caption{The amplitude of model power spectra fitted to QSO power
  spectra in redshift bins, assuming either EdS or $\Lambda$. In both
  cases, the shape parameter, $\Gamma$, was fixed at the best fitting
  value for that cosmology.}
\begin{center}
\begin{tabular}{@{}cccc}
\hline
Cosmology&Redshift bin&$\Gamma$&log(P(k=0.1))\\
\hline
EdS&$0.3<z<2.2$&0.16&$3.66\pm0.03$\\
EdS&$0.3<z<1.4$&0.16&$3.65\pm0.05$\\
EdS&$1.4<z<2.2$&0.16&$3.67\pm0.05$\\
$\Lambda$&$0.3<z<2.2$&0.13&$3.90\pm0.03$\\
$\Lambda$&$0.3<z<1.4$&0.13&$3.88\pm0.04$\\
$\Lambda$&$1.4<z<2.2$&0.13&$3.93\pm0.04$\\
\hline
\end{tabular}
\end{center}

\label{table}
\end{table}

\section{Comparison with models of large scale structure}
\label{sec:qsoimp}

\subsection{The {\it Hubble Volume} simulation}

\begin{figure} 
  \vspace{-2cm} \centerline{\hbox{\psfig{figure=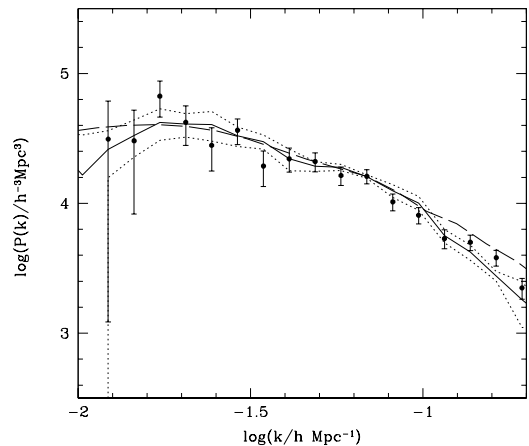,width=8.5cm}}}
\caption
{ The points show the 2QZ power spectrum with $\Lambda$ assumed. The
  dashed line shows the real space input power spectrum for the {\it
    Hubble Volume} simulation.  The solid line shows the {\it Hubble
    Volume} $\Lambda$CDM mock QSO catalogue redshift space power
  spectrum (see Hoyle et al. 2002), updated to match the final 2QZ
  selection function , with 1$\sigma$ errors indicated by the dotted
  lines.  }
\label{fig:pkhv}
\end{figure}

The {\it Hubble Volume} simulation was introduced in
section~\ref{hubvol}. The power spectrum determined from biased mock
QSO catalogues, created using the final 2QZ selection function
imprinted on the {\it Hubble Volume} simulation, is shown in
Fig.~\ref{fig:pkhv}, where it is compared to the real space input mass
power spectrum for the simulation, together with the 2QZ power spectrum
with $\Lambda$ assumed. The simulation was designed to mimic the 2QZ as
closely as possible, and so any differences in shape should give a good
test of the cosmological assumptions.

The first thing to note is that over a wide range of scales, the output
redshift-space mock catalogue power spectrum follows very closely the
real space input mass power spectrum for the simulation. On the largest
scales ($\log (k / h \,{\rm Mpc}^{-1})\la-1.5$), this demonstrates that
the effects of the 2QZ window, which were also imprinted on the {\it
  Hubble Volume} simulation, are relatively small, as shown in
section~\ref{wfsec}. This also shows that at scales of $\log (k / h
\,{\rm Mpc}^{-1})\la-1.0$ the biasing prescription we chose to create
the mock catalogues (see Hoyle et al. (2002) for details) has no effect
on the shape of the power spectrum. On smaller scales, however, we see
a slight reduction in the output power from the simulation, relative to
the real space input power spectrum. This can be attributed to
redshift-space distortions caused by small-scale peculiar velocities of
the mock QSOs.

A comparison between the {\it Hubble Volume} and 2QZ power spectra
shows that the simulation provides a very good fit to the observed data
over virtually the whole range observed. The 2QZ power spectrum is
therefore entirely consistent with the $\Lambda$CDM cosmological model
assumed to create the {\it Hubble Volume}.  Whilst performing detailed
analyses of N-body simulations such as the {\it Hubble Volume} is very
useful, providing a good test of the power spectrum estimator and a way
of examining possible systematic effects, it is also very time
consuming. We need to compare many different CDM models to derive
constraints on cosmological parameters. Therefore in the next section
we instead compare the QSO power spectrum to model CDM power spectra
with a variety of different cosmologies.

\subsection{Fitting model power spectra}

\begin{figure} 
  \vspace{-2cm}
  \centerline{\hbox{\psfig{figure=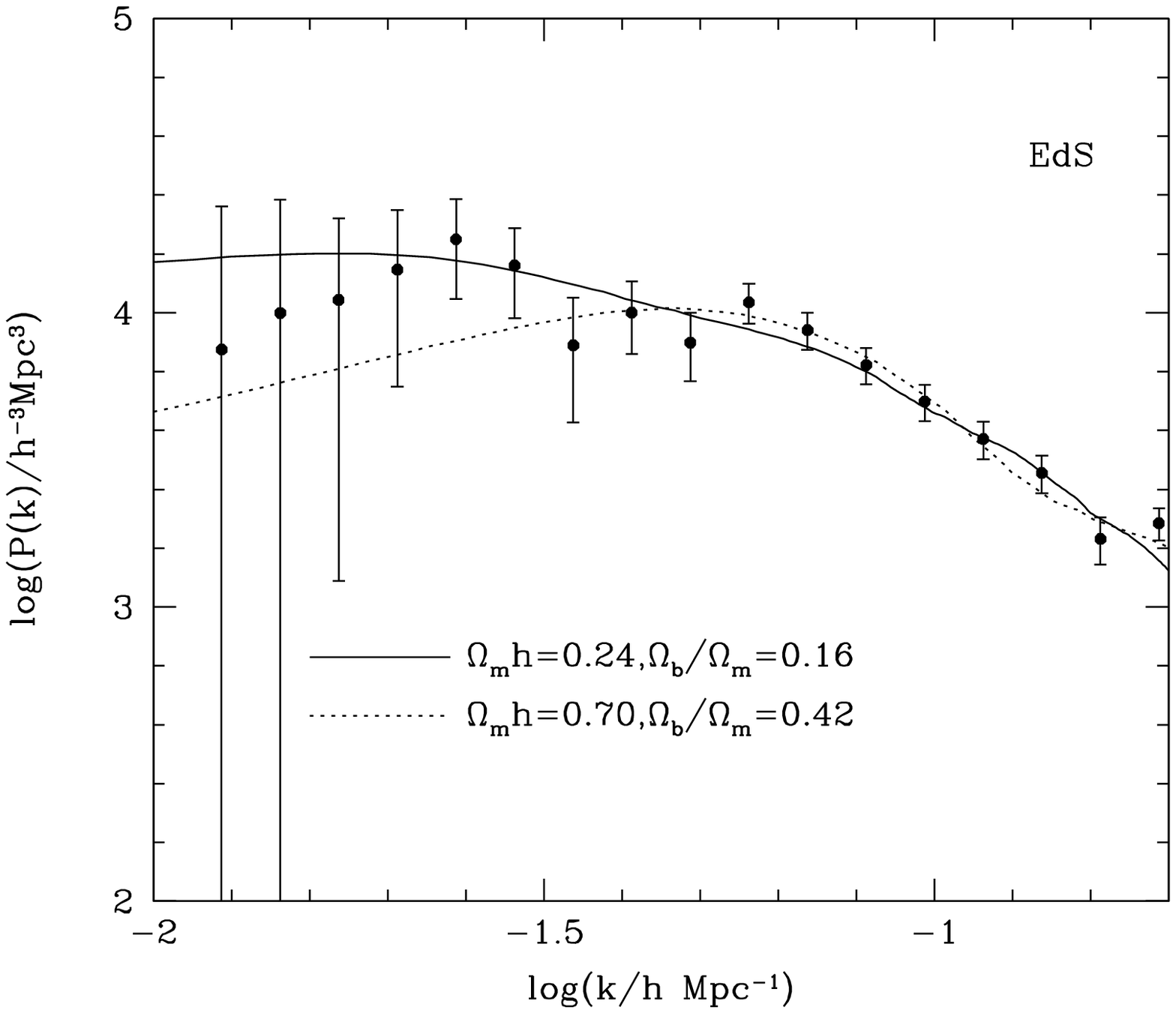,width=8.5cm}}}
  \vspace{-2cm}
  \centerline{\hbox{\psfig{figure=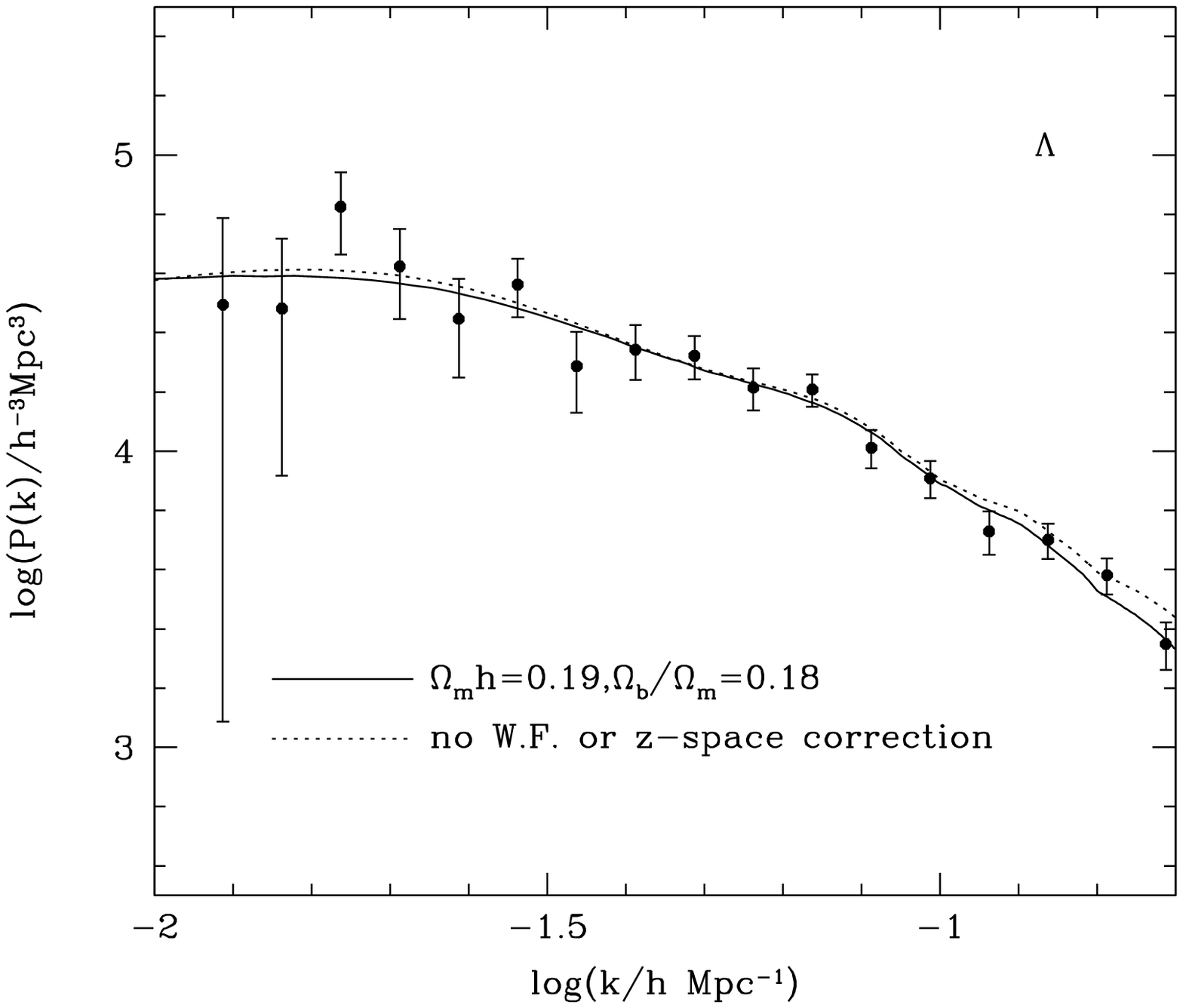,width=8.5cm}}}
\caption
{The plots show the best fit CDM models (solid lines) to the 2QZ power
  spectra (points). In the upper plot, the EdS $r(z)$ is assumed. The
  solid line shows the best fit CDM model with $\Omega_b/\Omega_m=0.18$
  and $\Omega_mh=0.24$. The dotted line shows a CDM model from the
  second branch of solutions ($\Omega_b/\Omega_m=0.42$ and
  $\Omega_mh=0.70$) that fit the 2dFGRS power spectrum.  The lower plot
  shows the best fit CDM model to the 2QZ power spectra assuming the
  $\Lambda$ $r(z)$, with $\Omega_b/\Omega_m=0.18$ and $\Omega_mh=0.19$
  (solid line). The dotted line shows, for comparison, the same model
  prior to correction for window function and redshift-space distortion
  effects.  }
\label{fig:pkmodel}
\end{figure}

\begin{figure} 
  \vspace{-0.5cm}
  \centerline{\hbox{\psfig{figure=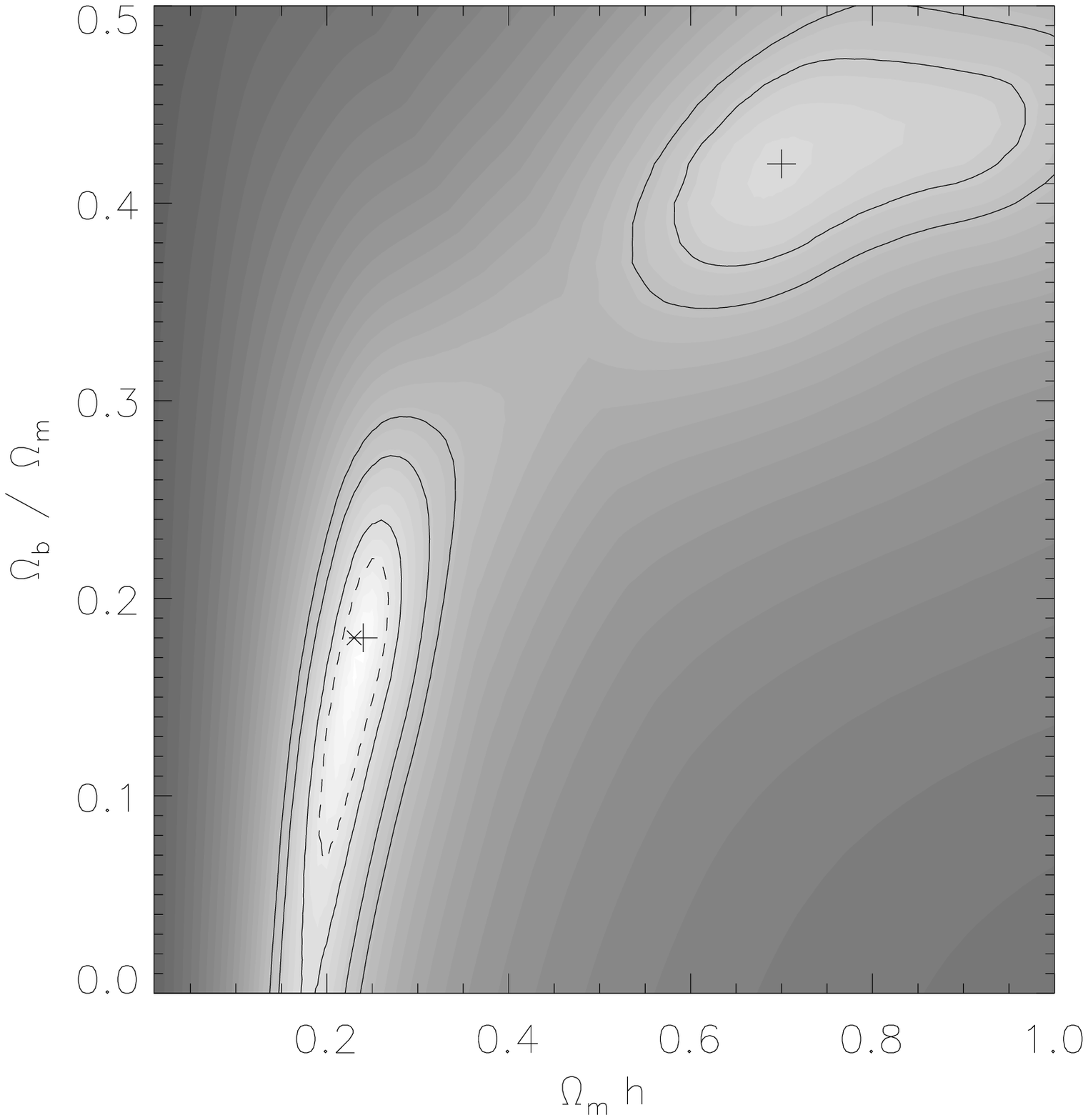,width=8cm}}}
  \vspace{-0.5cm}
  \centerline{\hbox{\psfig{figure=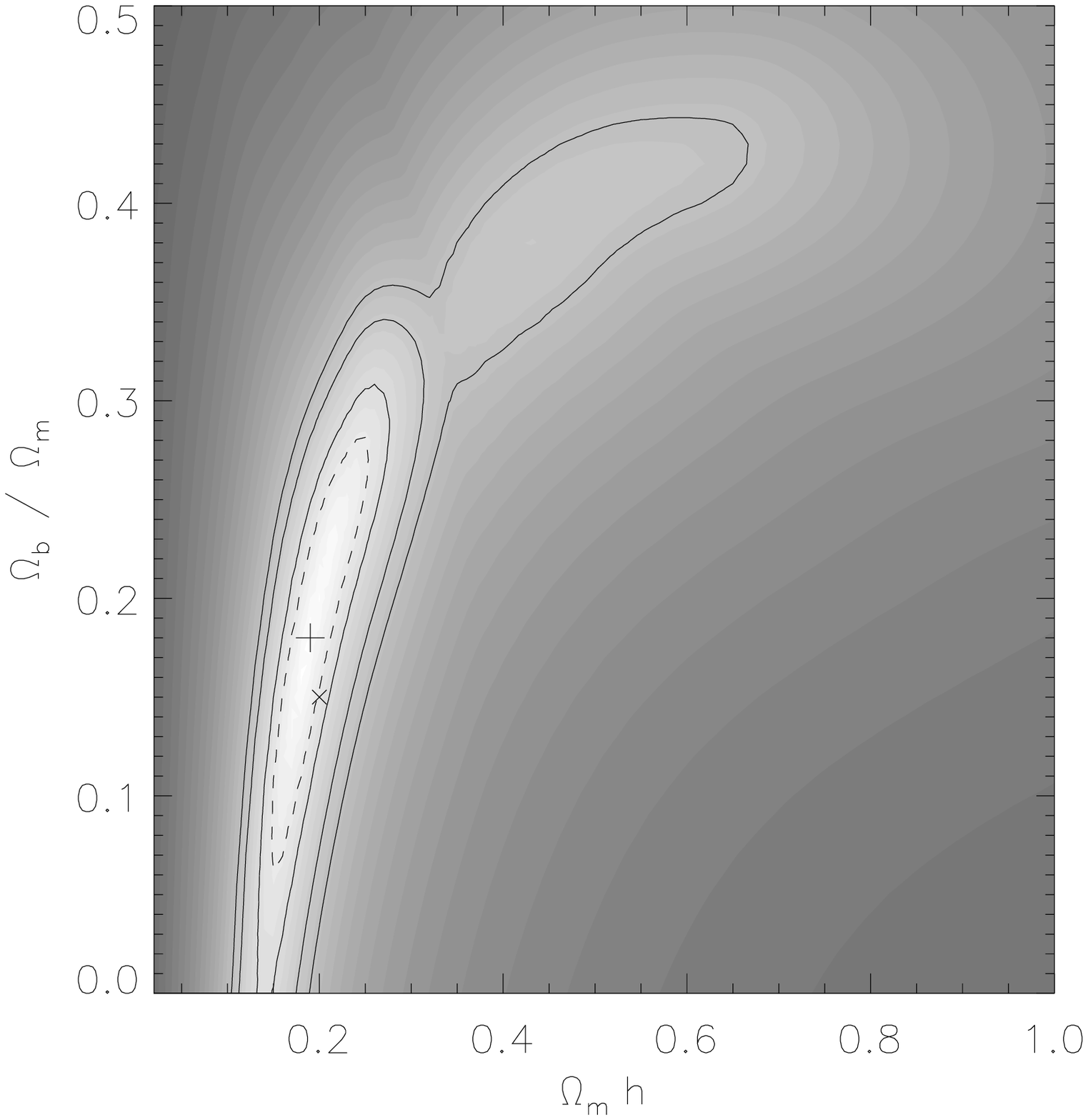,width=8cm}}}
\caption
{Filled contours of decreasing likelihood in the $\Omega_mh$ --
  $\Omega_b/\Omega_m$ plane, marginalizing over $h$ and the power
  spectrum amplitude, for fits to the 2QZ power spectra, assuming EdS
  for the upper plot, and $\Lambda$ for the lower plot. Countours are
  plotted for a one-parameter confidence of 68 per cent (dashed
  contour), and two-parameter confidence of 68, 95 and 99 per cent
  (solid contours). $+$ marks the best fit models to the 2QZ data
  ($\Omega_b/\Omega_m=0.18$, $\Omega_mh=0.24$, or with lower
  likelihood, $\Omega_b/\Omega_m=0.42$, $\Omega_mh=0.70$ assuming EdS,
  and $\Omega_b/\Omega_m=0.18$, $\Omega_mh=0.19$ assuming $\Lambda$),
  and $\times$ marks the best fit models ($\Omega_b/\Omega_m=0.18$,
  $\Omega_mh=0.23$ assuming EdS, and $\Omega_b/\Omega_m=0.15$,
  $\Omega_mh=0.2$ assuming $\Lambda$) determined in a similar analysis
  on the 2dFGRS data (Percival et al. 2001).  }
\label{fig:pkconstraint}
\end{figure}

\begin{figure} 
  \vspace{-0.5cm}
  \centerline{\hbox{\psfig{figure=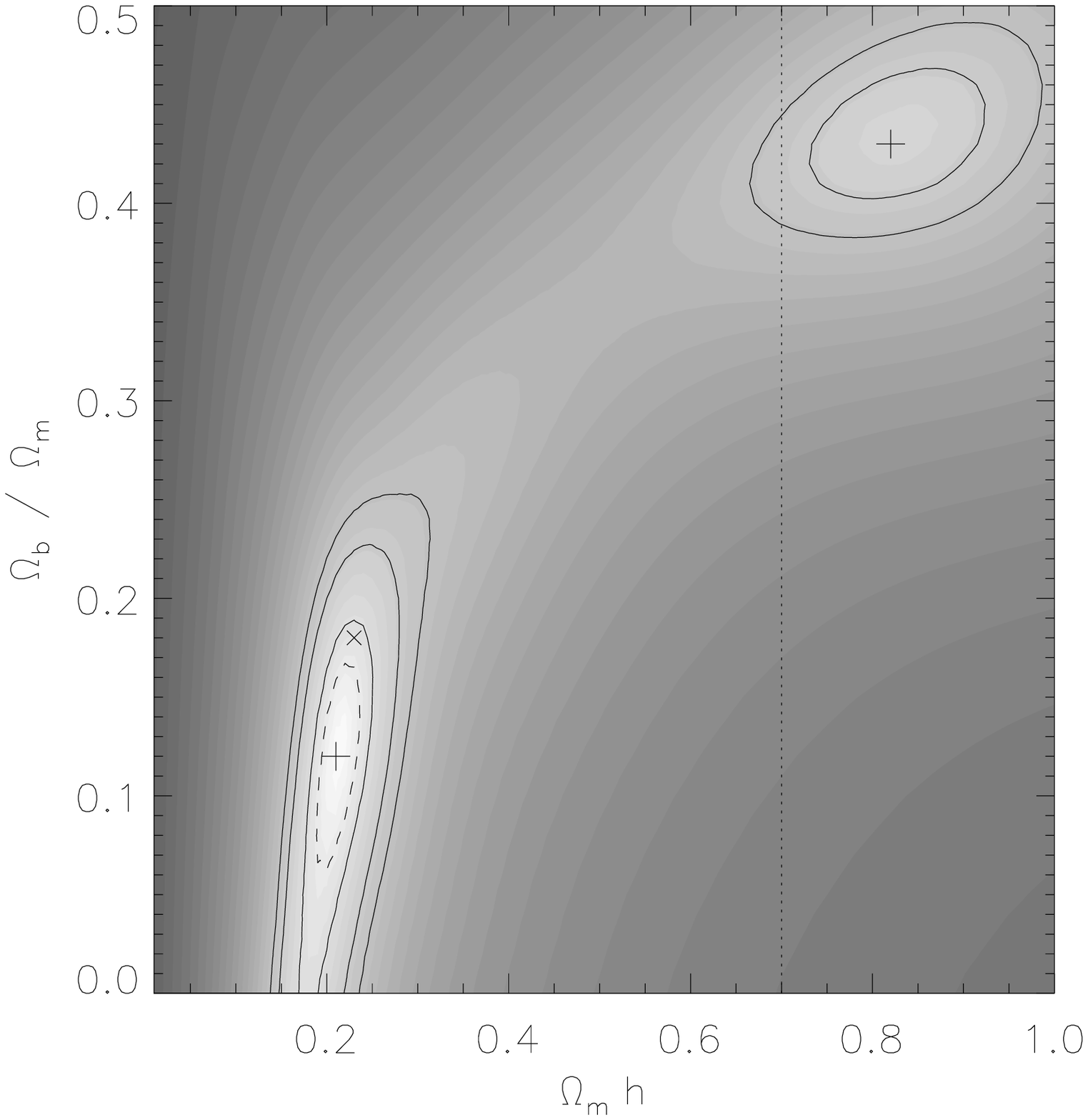,width=8cm}}}
  \vspace{-0.5cm}
  \centerline{\hbox{\psfig{figure=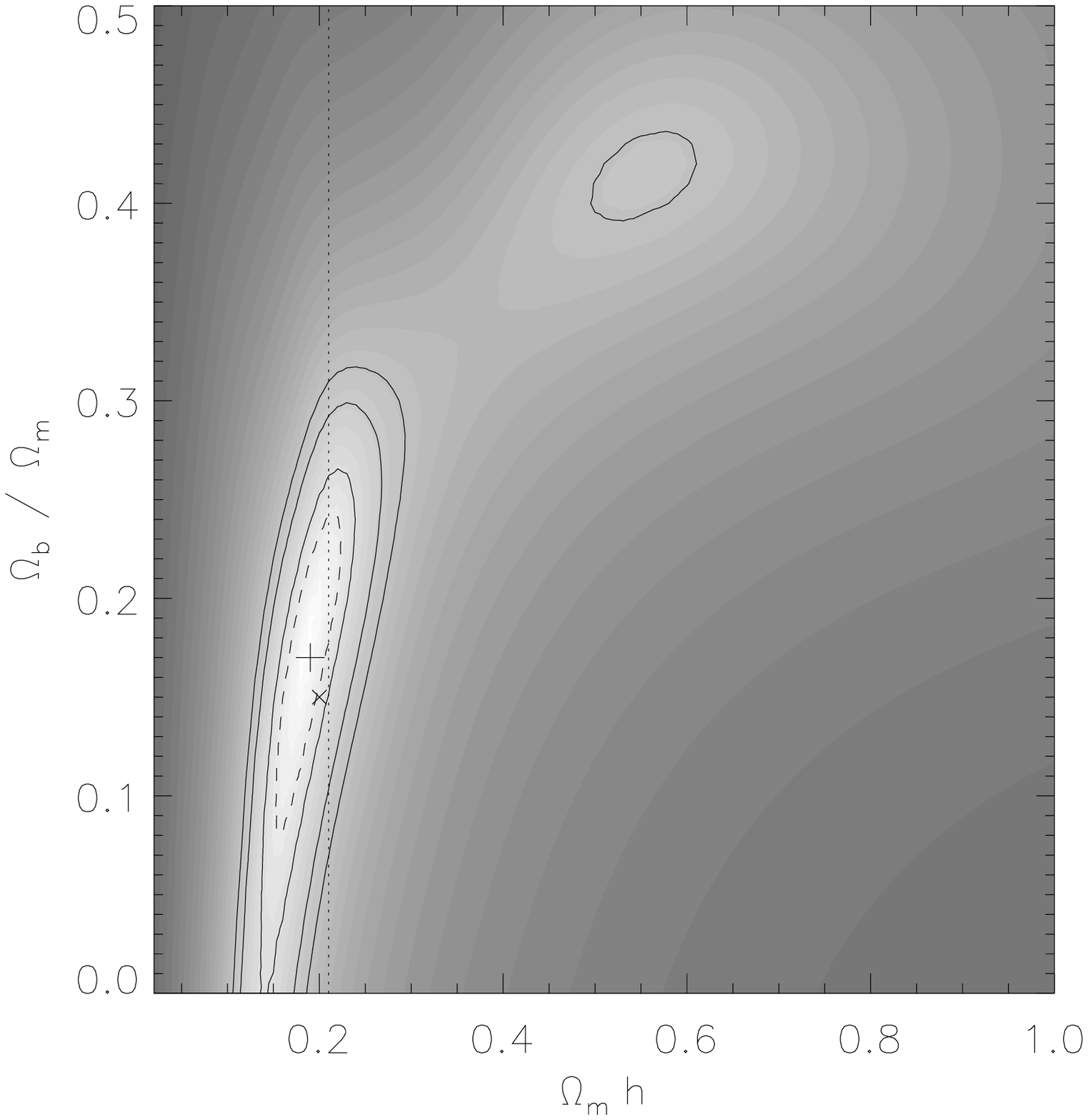,width=8cm}}}
\caption
{Filled contours of decreasing likelihood in the $\Omega_mh$ --
  $\Omega_b/\Omega_m$ plane, marginalizing the power spectrum amplitude
  but fixing $h=0.7$, for fits to the 2QZ power spectra, assuming EdS
  for the upper plot, and $\Lambda$ for the lower plot. Contours are
  plotted for a one-parameter confidence of 68 per cent (dashed
  contour), and two-parameter confidence of 68, 95 and 99 per cent
  (solid contours). $+$ marks the best fit model to the 2QZ data
  ($\Omega_b/\Omega_m=0.12$, $\Omega_mh=0.21$, or with lower
  likelihood, $\Omega_b/\Omega_m=0.43$, $\Omega_mh=0.82$ assuming EdS,
  and $\Omega_b/\Omega_m=0.17$, $\Omega_mh=0.19$ assuming $\Lambda$),
  and $\times$ marks the best fit models ($\Omega_b/\Omega_m=0.18$,
  $\Omega_mh=0.23$ assuming EdS, and $\Omega_b/\Omega_m=0.15$,
  $\Omega_mh=0.2$ assuming $\Lambda$) determined in a similar analysis
  on the 2dFGRS data (Percival et al. 2001). In each case, the dotted
  vertical line shows the value of $\Omega_mh$ assumed to calculate
  $r(z)$.  }
\label{fig:pkconstraint2}
\end{figure}

Assuming that QSO bias is not scale-dependent on the scales we are
probing, the shape of the QSO power spectrum provides new constraints
on the matter and baryonic contents of the Universe at a different
(much higher) redshift than that of the 2dFGRS, or other galaxy /
cluster power spectrum determinations. Therefore, the QSO power
spectrum measurement at $z\sim1.4$ can provide a powerful new test for
cosmological models that have previously only been constrained locally
($z\sim0$) and at recombination ($z\sim 1000$). Model power spectra
with a variety of CDM cosmologies have been created using transfer
function formulae from Eisenstein \& Hu (1998) to compare with the 2QZ
data. We have chosen to vary $\Omega_b/\Omega_m$ (where
$\Omega_m=\Omega_{cdm}+\Omega_b$), and $\Omega_mh$, considering the
range $0<\Omega_b/\Omega_m<0.5$ and $0<\Omega_mh<1.0$ in order to
determine constraints on these important parameters. We have assumed a
scale invariant spectrum of initial fluctuations.

The models generated have been numerically convolved with the 2QZ
window function. We noted in the comparison with the {\it Hubble
  Volume} simulation that small-scale power is slightly suppressed in
the redshift-space power spectrum by small-scale peculiar velocities of
the QSOs.  A second similar effect also affects the 2QZ sample. There
is an uncertainty in determining QSO redshifts from the low resolution,
low S/N 2dF spectra of $\delta z \sim 0.0035$ (Croom et al. 2001a),
which, at the average redshift of the survey, assuming $\Lambda$,
corresponds to an uncertainty in the line-of-sight distance of $\sim
5\;h^{-1}\;$Mpc. This introduces an apparent velocity dispersion of
$\sigma_p \sim 600\;$km$\;$s$^{-1}$, which can be added in quadrature
to the true QSO small-scale velocity dispersion. We can apply a simple
model to correct the power spectrum for these redshift-space
distortions (Ballinger et al. 1996). Following Outram et al. (2001) we
assume an apparent QSO line of sight pairwise velocity dispersion, due
to both redshift determination errors and peculiar velocities, of
$700\;$km$\;$s$^{-1}$, and a value of $\beta \approx \Omega_m^{0.6}/b =
0.39$ to describe the coherent peculiar velocities on large, linear
scales due to the infall of galaxies into overdense regions. We do not
attempt a correction for geometric distortions introduced by incorrect
cosmological model assumptions when measuring the power spectrum.

For this analysis, we have assumed that the data points are
independent. Due to the compactness of our window function, this should
be a good approximation. We have also had to assume a cosmology to
derive the QSO $r(z)$ prior to measuring the QSO power spectrum.
Ideally one would like to fit each power spectrum model to the power
spectrum measured assuming the appropriate cosmology. However, this
would be very time consuming, and beyond the scope of this paper.
Therefore we limit our analysis to power spectra measured assuming
either an EdS or $\Lambda$ $r(z)$. Despite assuming a particular
cosmology to derive the $r(z)$, we do not initially restrict the fitted
parameters to match these cosmologies. We determine the likelihood of
each model by calculating the $\chi^2$ value of each fit, performed
over the range $-1.9<\log(k/h\, {\rm Mpc}^{-1})<-0.75$, and assuming
that ${\cal L}\propto\exp(-\chi^2/2)$.  For each model, we allowed the
amplitude to vary (to account for linear QSO bias; we assumed that QSO
bias is scale independent on scales $\log(k/h\, {\rm
  Mpc}^{-1})<-0.75$), choosing the value that maximised the likelihood
of the fit.

We also allowed $h$ to vary, within the range $0.4<h<0.9$, choosing the
value that maximised the likelihood of the fit. As the shape of the
power spectrum is primarily dependent on $\Omega_mh$, and relatively
invariant to changes in $h$ for a given $\Omega_mh$, the value of $h$
has relatively little effect on the results. Repeating the fit, with
$h$ fixed at $h=0.7$, following, for example, the results of Tanvir et
al. (1995) and the HST Key Project (Mould et al. 2000) only marginally
improved the constraints on the parameters of interest.

The best fit models, shown in Fig.~\ref{fig:pkmodel}, fit the data well
over the full range of scales considered. The effect of the window
function and redshift-space corrections can be seen in the $\Lambda$
plot. The solid line in the plot shows the best fit model to the
$\Lambda$ 2QZ power spectrum, with $\Omega_b/\Omega_m=0.18$ and
$\Omega_mh=0.19$. The dotted line shows the same model prior to
correction for window function and redshift-space distortion effects.
The effect of the window function can be seen on large scales, slightly
supressing the power, and on small scales the effect of redshift-space
distortions can be seen. This lack of small scale power steepens the
power spectrum, leading to a lower value of $\Gamma$. If not corrected
for, therefore, it would lead to an under-estimate of $\Omega_mh$. The
effect of the window function is slightly larger when an EdS $r(z)$ is
assumed.

Likelihood contours in the $\Omega_mh$ -- $\Omega_b/\Omega_m$ plane for
fits to the 2QZ power spectra are shown in Fig.~\ref{fig:pkconstraint}
(allowing $h$ to vary), and Fig.~\ref{fig:pkconstraint2} (with fixed
$h=0.7$). Assuming an EdS QSO $r(z)$, we find a degeneracy between the
two parameters. The best fit model has $\Omega_b/\Omega_m=0.18\pm0.09$
and $\Omega_mh=0.24\pm0.04$, however a second solution with
$\Omega_b/\Omega_m=0.42$ and $\Omega_mh=0.70$ cannot be rejected at 95
per cent confidence from this analysis alone. For comparison, we show
the best fit derived from the 2dFGRS data (Percival et al. 2001), which
is in good agreement with the result derived here. The degeneracy we
see is also very much like that found in the 2dFGRS data.  We note that
the best fit value of $\Omega_mh$ obtained is totally inconsistent with
the assumed EdS cosmology. Whilst the second solution is consistent
with EdS, it is strongly rejected by other analyses such as estimates
of the baryon content from big-bang nucleosynthesis (O'Meara et al.
2001) and recent Cosmic Microwave Background (CMB) results (e.g.
Sievers et al. 2002).

The best fit model to the $\Lambda$ $r(z)$ 2QZ data has
$\Omega_b/\Omega_m=0.18\pm0.10$ and $\Omega_mh=0.19\pm0.05$. Although
the contours are again elongated in the direction of the degeneracy
seen in the EdS plot, the second peak seen in the greyscale is ruled
out at over 95 per cent confidence. Due to the huge volume of the 2QZ,
particularly in a $\Lambda$ cosmology, we are able probe larger scales
than with the 2dFGRS. Whilst the second solution is not a bad fit on
scales $\log(k/h\, {\rm Mpc}^{-1})>-1.5$, we can discriminate between
the two models on the largest scales probed by the 2QZ. When repeating
the fits, considering only models with $h=0.7$, we see a marginal
tightening of the contours, shown in Fig.~\ref{fig:pkconstraint2}, with
best fit models of $\Omega_b/\Omega_m=0.17\pm0.09$ and
$\Omega_mh=0.19\pm0.04$. This is totally consistent with the $\Lambda$
cosmology assumed to calculate $r(z)$.

These results agree well with those obtained by Yamamoto (2002) who
produced a similar analysis of the cosmological parameter constraints
that could be derived from the 2QZ 10k catalogue. The values of
$\Omega_mh$ and $\Omega_b/\Omega_m$ obtained assuming either $r(z)$ are
very similar, indicating that these results are largely independent of
our $r(z)$ assumption.  When fitting models in the $\Omega_mh$ --
$\Omega_b/\Omega_m$ plane we have considered only a family of CDM
models with a scale invariant spectrum of initial fluctuations ($n=1$).
If we relaxed these assumptions we would introduce new degeneracies
(Efstathiou et al. 2002), and significantly reduce the constraints on
the above parameters. However, the degeneracies between cosmological
parameters derived from power spectral analyses of large-scale
structure (e.g. probed by the 2QZ or the 2dFGRS) alone are different
from those from other datasets, such as the CMB, and Efstathiou et al.
demonstrated that by combining such datasets, these degeneracies can be
broken, leading to strong cosmological constraints.

\section{Conclusions}
\label{sec:qsoconc}

We have presented a power spectrum analysis of the final 2dF QSO
Redshift Survey catalogue containing 22652 QSOs. Utilising the huge
volume probed by the QSOs, we have accurately measured the power over
the range $-2<\log(k/h\, {\rm Mpc}^{-1})<-0.7$, thus covering over a
decade in scale, and reaching scales of $\sim 500h^{-1}$Mpc. The
$\Lambda$ $r(z)$ QSO power spectrum can be well described by a model
with shape parameter $\Gamma=0.13\pm0.02$. If an Einstein-de Sitter
model is instead assumed, a slightly higher value of
$\Gamma=0.16\pm0.03$ is obtained. These low measurements of $\Gamma$
indicate significant large-scale power and assuming either cosmology we
can strongly rule out a turnover in the power spectrum at scales below
$\sim$300$h^{-1}$Mpc ($\log (k / h \,{\rm Mpc}^{-1})\sim-1.7$).  The
amplitude of clustering of the QSOs at $z\sim1.4$ is similar to that of
present day galaxies, and an order of magnitude lower than present day
clusters. A comparison of the power spectra of QSOs at high and low
redshift shows little evidence for any evolution in the amplitude of
clustering.

A comparison with the {\it Hubble Volume} $\Lambda$CDM simulation shows
very good agreement over the whole range of scales considered. To
derive constraints on the matter and baryonic contents of the Universe,
we fit CDM model power spectra (assuming scale-independent bias and
scale-invariant initial fluctuations), convolved with the survey window
function, and corrected for redshift space distortions. Assuming a
$\Lambda$ cosmology $r(z)$, we find that models with baryon
oscillations are slightly preferred, with the baryon fraction
$\Omega_b/\Omega_m=0.18\pm0.10$. The overall shape of the power
spectrum provides a strong constraint on $\Omega_mh$ (where $h$ is the
Hubble parameter), with $\Omega_mh=0.19\pm0.05$. These new constraints
are derived at $z\sim1.4$, which corresponds to a look-back time of
approximately two-thirds of the age of the Universe. Cosmological
constraints at this intermediate epoch provide a strong test for models
previously constrained to match CMB observations at $z\sim1000$ and the
local Universe at $z\sim0$.

\section*{Acknowledgements} 
The 2dF QSO Redshift Survey was based on observations made with the
Anglo-Australian Telescope and the UK Schmidt Telescope, and we would
like to thank our colleagues on the 2dF Galaxy Redshift Survey team and
all the staff at the AAT that have helped to make this survey possible.
We would like to thank Adrian Jenkins, Gus Evrard, and the Virgo
Consortium for kindly providing the {\it Hubble Volume} simulations.
PJO would like to acknowledge the support of a PPARC Fellowship.

\end{document}